\documentstyle[preprint,aps,epsfig]{revtex} 

\renewcommand{\today}{25th July 2000}

\newcommand{\nc}{\newcommand} 

\nc{\gam}{\gamma \gamma} 
\nc{\be}{\begin{equation}}
\nc{\ee}{\end{equation}} 
\nc{\bea}{\begin{eqnarray}}
\nc{\eea}{\end{eqnarray}} \nc{\beas}{\begin{eqnarray*}}
\nc{\eeas}{\end{eqnarray*}} \nc{\noi}{\noindent} \nc{\sD}{\not \! \!
D} \nc{\s}[1]{\not \! #1} \nc{\non}{\nonumber} \nc{\bb}{\bibitem}
\nc{\lf}{\left} \nc{\ri}{\right} \nc{\mb}[1]{\makebox[#1]{}}
\nc{\pa}{\partial} \nc{\sA}{\not \! \! A}
\nc{\newsec}[1]{\section{#1}\mb{0.5cm}} \nc{\h}{\frac{1}{2}}
\nc{\ra}{\rightarrow} \nc{\la}{\leftarrow}
\nc{\ep}{$e^+e^-\ra\pi^+\pi^-\;$} \nc{\emuon}{$e^+e^-\ra\mu^+\mu^-\;$}
\nc{\epp}{$e^+e^-\ra\pi^+\pi^0\pi^-\;$}
\nc{\elec}{$e^+e^-\ra\gamma^*\ra e^+e^-\;$}
\def\mathunderaccent#1{\let\theaccent#1\mathpalette\putaccentunder}
\def\putaccentunder#1#2{\oalign{$#1#2$\crcr\hidewidth \vbox
to.2ex{\hbox{$#1\theaccent{}$}\vss}\hidewidth}}
 \nc{\ti}{\mathunderaccent\tilde}
\nc{\M}{{\cal M}} \nc{\rw}{$\rho\!-\!\omega\;$} \def\hhht{\rule[
0.mm]{0.mm}{6.mm}} 
\def\hhhb{\rule[-3.mm]{0.mm}{9.mm}}
\def\hhhc{\rule[-3.mm]{0.mm}{3.mm}}
\def\hhhd{\rule[-3.mm]{0.mm}{1.mm}}
\def\hhhu{\rule[-3.mm]{0.mm}{12.mm}}

\begin{document}
\tightenlines 
\preprint{\vbox{ Eur.\ Phys.\ J.\ C {\bf 17}, 593 (2000) 
\hfill LPNHE 99--04 \\ \null \hfill SLAC--PUB--8097 \\ \null
\hfill hep-ph/9905350}}

\title{VMD, the WZW Lagrangian and ChPT:\\[0.5cm] The Third Mixing
Angle}

\author{M.~Benayoun, L. DelBuono} \address{LPNHE des Universit\'es
Paris VI et VII--IN2P3, Paris, France}

\author{H.B. O'Connell\thanks{Work supported by the US Department of
Energy contract DE-AC03-76SF00515.}}  \address{Stanford Linear
Accelerator Center, Stanford University, Stanford CA 94309, USA}

\date {Preliminary Release 14th May 1999, Revised \today}

\maketitle
\begin{abstract}

We show that the Hidden Local Symmetry Model, supplemented with 
well-known
procedures for breaking flavor SU(3) and nonet symmetry, provides all
the information contained in the standard Chiral Perturbation Theory
(ChPT) Lagrangian ${\cal L}^{(0)}+{\cal L}^{(1)}$. This allows to
rely on  radiative decays of light mesons ($VP\gamma$ and
$P \gamma\gamma$) in order to extract some numerical information of 
relevance to ChPT: a
value for $\Lambda_1=0.20 \pm 0.04$, a quark mass ratio of $\simeq
21.2 \pm 2.4$, and a negligible departure from the Gell-Mann--Okubo
mass formula. The mixing angles are $\theta_8=-20.40^\circ \pm
0.96^\circ$ and $\theta_0=-0.05^\circ \pm 0.99^\circ$. We also give
the values of all decay constants.  It is shown that the common mixing
pattern with one mixing angle $\theta_P$ is actually quite appropriate
and algebraically related to the $\eta/\eta'$ mixing pattern presently
preferred by the ChPT community.  For instance the traditional $\theta_P$ is
functionally related to the ChPT $\theta_8$ and fulfills $\theta_P
\simeq \theta_8/2$. The vanishing of $\theta_0$, supported by all data
on radiative decays, gives a novel relation between mixing angles and
the violation of nonet symmetry in the pseudoscalar sector. Finally,
it is shown that the interplay of nonet symmetry breaking through U(3)
$\ra$ SU(3)$\times$ U(1) satisfies all requirements of the physics of
radiative decays without any need for additional glueballs.
\end{abstract}

\newpage

\pagenumbering{arabic}
\section{Introduction}

We have recently proposed a model for radiative decays of all light
mesons \cite{rad} which gives a consistent and successful description
of all reported experimental information. This covers the 14 decay
modes of the kind{\footnote{ This counting does not include $\pi^0 \ra
\gamma \gamma$, which would serve to fix $f_\pi$. We prefered using
directly the PDG recommended value \cite{PDG98} as for $f_K$.}}  $V
\rightarrow P \gamma$ and $P \rightarrow \gamma \gamma$. This vector
meson dominance (VMD) based model relies on the hidden local symmetry
(HLS) approach developed in Ref.~\cite{HLS} which introduces the
vector mesons as gauge bosons of a spontaneously broken hidden local
symmetry and closely resembles Seiberg's EM duality in supersymmetric
QCD~\cite{Seiberg:1995pq}, as noted in Ref.~\cite{heath}
and again
in \cite{Harada:1999zj}.  Its anomalous sector \cite{HLS,FKTUY}
(referred to hereafter as FKTUY), describes the  radiative
decays  of light flavor mesons. In its original form this Lagrangian
is  U(3) symmetric, as it possesses both nonet symmetry 
and SU(3) flavor symmetry. 

In order to describe the full pattern of light
mesons radiative decays, these unbroken schemes need to be 
supplemented with symmetry breaking mechanisms.
Breaking the SU(3) flavor symmetry is an essential
step \cite{rad}. This is performed following 
the mechanism proposed by Bando, Kugo and Yamawaki (BKY)
\cite{BKY,heath} and does not depend on any additional free 
parameter.  An additional breaking procedure \cite{rad,conf,mixing} is needed 
in order  to describe the 
observed \cite{PDG98} features of $K^*$ radiative decays;
it allows us to recover for this sector a structure derived by  G. Morpurgo
in his approach to low energy QCD \cite{morpurgo}. 

An explicit form \cite{odonnel} of nonet symmetry breaking (NSB) for vector (V) 
mesons seems to play a negligible role \cite{mixing} when  focusing on  radiative 
decays. Angular departures from ideal mixing are instead highly significant;
they can be essentially explained  by the $\omega_I /\phi_I$ transitions{\footnote{
The subscript $I$  indicates the ideal combinations.}} inherent to SU(3) VMD models 
like HLS. One cannot, however,  completely exclude that some kind of 
vector NSB is hidden inside these angular effects \cite{mixing}.

Instead, NSB for pseudoscalar mesons (PS) is an essential ingredient \cite{rad}. 
It has been performed in the manner of Ref.~\cite{odonnel}, which turns out to 
allow couplings to singlet and octet PS components to be different.
It is a purpose of the present paper to revisit the 
issue of how NSB can be consistently implemented within the HLS Lagrangian.

\hspace{1.cm}

The  problem of $\eta/\eta'$ mixing \cite{DHL,GILMAN} is tightly
connected with the breaking of nonet symmetry.
This can be performed at the level of the coupling constants
\cite{odonnel}, but this breaking can also be connected 
with a possible glue component inside light mesons 
\cite{ball,veneziano,basu1,basu2}. Indeed, the full set of radiative
decays \cite{rad} examplify the large importance of this effect.
Within this context, this reference also showed that effects 
of such a glue component  can only affect the $\eta'$ meson, but 
cannot be disentangled from genuine nonet symmetry breaking effects
without some {\it a priori} knowledge of one of these twin phenomena. 

In connection with this particular problem, but even more 
closely related with the effects of symmetry breakdown in Chiral 
Perturbation Theory (ChPT), Kaiser and Leutwyler \cite{leutw,leutwb}
advocate an $\eta/\eta'$ mixing scheme (see Ref. \cite{feldmann} for a 
comprehensive review), more complicated than the usual one, depending 
on two decay constants and  two mixing angles.
Some phenomenological analyses \cite{kroll1,kroll2} have  
investigated this new scheme. In a more axiomatic approach to QCD,
Shore \cite{shore} also finds appropriate a four parameter parametrization
of the $\eta/\eta'$ mixing.

\hspace{1.cm}

Nevertheless, the analysis of radiative 
decays (however, 14 independent decay modes)
of  Ref.~\cite{rad} does not find any need for  a 
four--parameter structure of the $\eta/\eta'$ mixing, 
as if phenomenology were exhibiting several relations among the 4 
decay constants \cite{leutw,leutwb,shore}, which might be fulfilled
at the (already high) level of accuracy permitted by the data.
To be more precise, Ref. \cite{rad} yields a quite
satisfactory description of the data by introducing
one PS mixing angle and one NSB parameter affecting
the PS sector; additional departures from SU(3)
flavor symmetry as per BKY \cite{BKY,heath} arise
only through a  dependence in $f_K/f_\pi$, which
can hardly be considered as a (free) parameter.  

More appealing, the anomalous Lagrangian
of Wess, Zumino and Witten (WZW) \cite{WZW1,WZW2}, 
with  SU(3) symmetry broken as explained in Refs.~\cite{rad}, 
leads to definitions of the mixing angle and decay constants 
as per Current Algebra and as following from the HLS--FKTUY framework. 
The (single) mixing angle  was found in Ref. \cite{rad} to be  $\simeq -10^{\circ}$, 
and,  moreover, the value for the octet decay constant is $f_8=0.82 f_{\pi}$.
The relevance of these parameter values is strongly supported by an
impressive agreement with experimental data within a highly
constrained model (5 parameters for 14 decay modes). 
Interestingly, Ref. \cite{ukqcd}, relying on lattice
QCD calculations reaches also a mixing angle a value 
$\simeq -10^\circ \pm 2^\circ$, with a preference
for $-10.2^\circ$.

On the other hand, there are repeated claims  \cite{scadron1,scadron1b,scadron2}
that a (single) mixing angle, as coming from standard Current Algebra
expressions, is quite appropriate and  is found to be much less
negative ($\simeq -13^\circ$ to $ \simeq -15^\circ$) than expected from 
ChPT ($\simeq -20^\circ$). A quite detailed discussion 
of this can be found in Ref. \cite{kekez} (see also  
\cite{scadron3}) where such
an angle value is derived from a bound state approach.
Another Lagrangian approach \cite{gedalin}, parent to HLS using
a specific breaking scheme,
recently claimed an angle value close to the previously
mentioned ones ($\simeq -15.4^\circ \pm 1.8^\circ$).

All this seems in glaring disagreement with the expectations of ChPT
\cite{DHL,GILMAN,leutw,leutwb,GL85,DGH}. Taking into account the special 
role of ChPT in low energy phenomenology, a possible contradiction between 
ChPT,  lattice QCD calculations or the VMD conceptual framework{\footnote{Or other approaches as listed
in the throughout discussion in Ref. \cite{kekez}.}} is a worrying question which must 
be  addressed and understood. This is the purpose of the present 
paper, which will show that the contradiction is illusory and only due to different
definitions of the same parameters in a naive understanding of the
WZW approach (with encompasses the Current Algebra definitions) and in ChPT.
We shall explicitly state the relationships between them. 

It will be shown that the VMD approach, relying on the HLS model 
broken as in Ref.~\cite{rad}, is actually in accord with all ChPT expectations
associated with the ChPT Lagrangian \cite{leutw,leutwb,feldmann} 
${\cal L}^{(0)}+ {\cal L}^{(1)}$. 
This will be illustrated by deriving from a broken VMD Lagrangian
model, all known leading order expressions for ChPT mixing
angles and decay constants, and by deriving their expected numerical values.
For the sake of  conciseness, we  shall frequently use NSB to refer
to nonet symmetry breaking and to FSB for SU(3) flavor symmetry breaking.


The outline of the paper is as follows. In Section \ref{VMDmodel}
we present a Lagrangian VMD model, based on the HLS approach, which includes
both NSB and FSB. We show that there is a close connection between them.
In Section~\ref{lagnsb}, we  study the field transformation
which permits us to write the kinetic energy of this broken VMD
model in canonical form, in terms of renormalized fields. We show
here that the field transformation of Refs. \cite{rad,mixing} corresponds
to a first order truncation in both NSB and FSB.

Section \ref{anomalous}
gives the VMD description of the $\eta/\eta' \rightarrow \gamma \gamma$
decay which depends on one mixing angle and nonet symmetry breaking 
(as parametrized by $x$). In 
Section \ref{wzwl1} the corresponding description is derived, starting
from the BKY broken WZW Lagrangian, and it is shown that WZW and VMD coincide.

In Section \ref{chpt}, we derive the set of relations which allows one
to define mixing angles and decay constants in accord with the standard
(or extended) ChPT approach. Here we show, first that
the definitions of mixing angles and decay constants from VMD/WZW 
and ChPT do not coincide once symmetry is broken
and, second, that VMD provides expressions and values for all
accessible  ChPT parameters in accord with expectations. 
This is illustrated by several examples, including the
functional  relation between the VMD mixing angle $\theta_P $ and
the ChPT angle $\theta_8$.

In Section \ref{comparison}, we show that starting from the
axial anomaly, it is possible to reconstruct the one angle
mixing scheme as it arises in our broken VMD model
and from the WZW Lagrangian; we comment on the previous use
of ChPT predictions in phenomenological analyses
of radiative decays data.

In Section \ref{feedback}, we show that nonet symmetry breaking
and pseudoscalar mixing angle(s) are functionally related,
which is a completely new result. This allows us to perform a fit 
of radiative decays with only 4 free parameters.
The level of nonet symmetry breaking correlated with the
fit value of the pseudoscalar mixing angle is shown
to remove any need for glue in the $\eta'$ meson.
A few other points of interest are also examined 
(quark mass ratio, isoscalar mass matrix, effects of NSB on 
PS mixing angle values).
Finally, Section \ref{conclud} is devoted to conclusions. 

\section{A Broken HLS Model for Radiative Decays}
\label{VMDmodel}  

\indent \indent The model developed in Ref. \cite{rad}
in order to describe all light meson radiative decays
relies on breaking nonet symmetry and flavor SU(3) in
the HLS Lagrangian \cite{HLS}, and especially  in its
anomalous (FKTUY) sector \cite{FKTUY}. The breaking
procedure of SU(3)  flavor symmetry (referred to hereafter as FSB)
in the non--anomalous HLS Lagrangian is the so--called new scheme, 
a variant the  original BKY  breaking mechanism \cite{BKY}
discussed in  Ref.  \cite{heath}.

For the purpose of only studying light meson radiative decays
\cite{rad}, a detailed knowledge of the nonet symmetry breaking (NSB)
mechanism is not needed; one only needs to know the field 
renormalization it would imply. The choice made in  Ref.~\cite{rad} 
was to postulate a likely form; this was determined by the O'Donnell 
derivation  of the SU(3) -- not U(3) -- $VP\gamma$ couplings, which assumes 
only the SU(3) flavor group structure, gauge invariance and Lorentz 
invariance \cite{odonnel}.

However, the way FSB and NSB in the PS sector merge together
is a much stronger assumption which only relies on its impressive 
phenomenological success \cite{rad} when describing the full set 
of radiative decays of light mesons. In this section, we aim at proposing a 
Lagrangian model which provides the appropriate  PS  field renormalization; it allows
to strongly motivate this assumption, by relating this Lagrangian to 
the ChPT framework.

\subsection{Basic Ingredients}

The basic ingredients of the effective Lagrangian
approach to the interaction of vector and pseudoscalar mesons
are the matrices $V$  and $P$ of the vector and pseudoscalar fields
expressed in the flavor ($u$, $d$, $s$) basis.
The vector meson field matrix $V$ is usually written{\footnote{
The sign in front of $\phi^I$ means that we define 
$\phi^I=-|s\overline{s}\rangle$.}} in terms of ideally mixed states 
($\omega^I$, $\phi^I$)
\be
V\equiv V^aT^a=\frac{1}{\sqrt{2}}
  \left( \begin{array}{ccc}
   (\rho^0+\omega^I)/\sqrt{2}  & \rho^+             &  K^{*+} \\
            \rho^-           & (-\rho^0+\omega^I)/\sqrt{2}    &  K^{*0} \\
            K^{*-}           & \overline{K}^{*0}  &  -\phi^I   \\
         \end{array}
  \right),\,\,\,a=1,...,8.
\label{vectfield}
\ee
Correspondingly, the pseudoscalar field matrix is usually defined as
\bea\non
P\equiv P^{a^\prime}T^{a^\prime}&=&\frac{1}{\sqrt{2}}
  \left( \begin{array}{ccc}
            \frac{1}{\sqrt{2}}\pi^0+\frac{1}{\sqrt{6}}\pi_8+
            \frac{1}{\sqrt{3}}\eta_0&\pi^+ &  K^+ \\[0.3cm]
            \pi^-  & -\frac{1}{\sqrt{2}}\pi^0+\frac{1}{\sqrt{6}}\pi_8
            +\frac{1}{\sqrt{3}}\eta_0  &  K^0 \\[0.3cm]
            K^-             &  \overline{K}^0  &
             -\sqrt{\frac{2}{3}}\pi_8 +\frac{1}{\sqrt{3}}\eta_0 \\[0.3cm]
         \end{array} 
  \right),\\
&&a^\prime=0,...,8
\label{psfield}  
\eea
using the conventional octet and singlet components
($\pi_8$, $\eta_0$) for the isoscalar mesons.
For definiteness, the SU(3) matrices will be denoted $T^a$ ($a=1, \cdots 8$)
and fulfill the normalization condition Tr$[T^aT^b]=\delta^{ab}/2$.
We complete this matrix basis, by adding the unit matrix suitably normalized
$T^0\equiv{\bf 1}/\sqrt{6}$. 

The physical states ($\omega$, $\phi$, $\eta$, $\eta'$) are generated from 
the ideally mixed states by means of standard rotation angles $\delta_V$ 
or $\delta_P$ for vector and pseudoscalar mesons. Correspondingly,
the rotation angles for the singlet and octet states to the physically
observed mesons are traditionally named $\theta_V$ and $\theta_P$.
These well known relations can be found in 
Refs.~\cite{rad,odonnel,heath,PDG98}. The connection between ideal
and physical $\omega$ and $\phi$ fields is treated heuristically
in Ref. \cite{heath} and rigorously in Ref. \cite{mixing}. We recall  
for further use the traditional (one angle) expression
\begin{equation}
\left[
     \begin{array}{ll}
     \displaystyle \eta   \\[0.5cm]
     \displaystyle \eta'   
     \end{array}
\right]
=
\left[
     \begin{array}{lll}
\displaystyle \cos{\theta_P} & -\displaystyle \sin{\theta_P} \\[0.5cm]
\displaystyle \sin{\theta_P} &
\displaystyle ~~\cos{\theta_P} 
     \end{array}
\right]
\left[
     \begin{array}{ll}
     \pi_8\\[0.5cm]
     \eta_0\\
     \end{array}
\right]
\label{psrot}
\end{equation}
If  fields undergo renormalization, the fields $\pi_8$ and $\eta_0$ 
in the expression above should be understood renormalized \cite{heath}.
With a slightly liberalized, but obvious, notation,
the expressions above for $V$ and $P$ can also be written
\begin{equation}
V=V_8+V_0~, \,\,\,\,\,
P=P_8+P_0 
\label{u3sym} 
\end{equation}
which  exhibit their octet and singlet component combinations,
and show that nonet (U(3)) symmetry is implicitly assumed.

\subsection{Physical Motivation for Nonet Symmetry Breaking (NSB)}

Referring to O'Donnell \cite{odonnel}, NSB implies modifying 
Eq.~(\ref{u3sym}) to
\begin{equation}
V=V_8+yV_0 , \,\,\,\,\,
P=P_8+xP_0 
\label{u3brk} 
\end{equation}
In this way, NSB  changes the relative weight of the octet 
and singlet parts in {\it a priori} both meson sectors.
Ref. \cite{mixing} has recently performed a throughout
study of NSB in the vector sector and clearly concluded
that data were consistent with no such NSB ({\it i.e.} $y=1$);
it is therefore a motivated choice to neglect vector  NSB
and state $y=1$ definitely.

Another way to account for nonet symmetry breaking is to assume
that the singlet sector contains a component other than the standard
SU(3)/U(3) singlet; we name it glue only for convenience.
A possible coupling of the $\eta/\eta'$ doublet to glue can
be accounted for \cite{rad} by means of an additional angle $\gamma$,
which is zero if one chooses to decouple this doublet from glue
\begin{equation}
\left[
     \begin{array}{ll}
     \displaystyle \eta   \\[0.5cm]
     \displaystyle \eta'  \\[0.5cm] 
     \displaystyle \eta''  
     \end{array}
\right]
=
\left[
     \begin{array}{lll}
\displaystyle \cos{\theta_P} & -\displaystyle \sin{\theta_P} &0\\[0.5cm]
\displaystyle \sin{\theta_P}\cos{\gamma} &
\displaystyle \cos{\theta_P}\cos{\gamma} & \sin{\gamma}\\[0.5cm]
\displaystyle -\sin{\theta_P} \sin{\gamma} &  -\cos{\theta_P} 
\sin{\gamma}&\cos{\gamma}\\
     \end{array}
\right]
\left[
     \begin{array}{ll}
     \pi_8\\[0.5cm]
     \eta_0\\[0.5cm]
     gg\\
     \end{array}
\right].
\label{mixing}
\end{equation}

Indeed, following the analysis of Ref.~\cite{rad}, we do not
have to introduce any coupling of the $\eta$ meson to glue, 
which would introduce an additional angle ($\beta$ in 
Ref.~\cite{rad}). The angle
$\gamma$ produces a coupling of (only) the $\eta'$ meson to glue.
We have named $\eta''$ the possible triplet companion of the 
$\eta/\eta'$ mesons, and do not attempt to identify it
{\footnote{We shall not also attempt to include this additional singlet 
in the Lagrangian model to be proposed for reasons which will become 
clear at the end of this paper.}}.
If $\gamma=0$, one clearly recovers the usual mixing pattern
for the $\eta/\eta'$ system by decoupling it from glue. 

When fitting the data on radiative decays of light mesons,
the level of correlation between $x$ and $\gamma$ is found such 
that assuming glue and exact nonet symmetry ($x=1$), or assuming no glue 
($\gamma=0$) and some NSB ($x\simeq 0.9$), provide  the same 
description of the data \cite{rad,conf}. Therefore, whether glue is required
in order to describe the $\eta'$ properties is still a pending question 
which will be addressed in the present paper (see Section \ref{feedback}),
when an educated guess about the value of $x$ will be made.
Let us note that the level of glue can be as large as  20\% if nonet
symmetry breaking is ignored \cite{rad,conf} by setting $x=1$~;
this conclusion has been reached also by others 
\cite{kou,basu1,basu2}.

Clearly, in order that  our conclusion on this point be of relevance,
the exact meaning of $x$ should be exhibited; a framework
as reliable  as ChPT is appropriate. This also motivates
our goal of comparing our broken HLS--FKTUY framework
to ChPT.

\subsection{Basics of the HLS Model}
\label{basics}

We refer the reader to Ref.~\cite{HLS} for a comprehensive
review of the  HLS model. A brief account can be found
in Ref.~\cite{heath}. We only recall the main features here.

The HLS Lagrangian can be written 
${\cal L}_{\rm HLS}={\cal L}_A + a {\cal L}_V$, where
\bea\non
{\cal L}_{A}&=&-\frac{f_{\pi}^2}{4}{\rm Tr}\left[
D_\mu\xi_L\xi_L^{\dagger}-D_\mu\xi_R\xi_R^{\dagger}\right]^2
\equiv-\frac{f_\pi^2}{4}{\rm Tr}[L-R]^2
\\
{\cal L}_V&=&-\frac{f_{\pi}^2}{4}{\rm Tr}\left[
D_\mu\xi_L\xi_L^{\dagger}+D_\mu\xi_R\xi_R^{\dagger}\right]^2
\equiv-\frac{f_{\pi}^2}{4}{\rm Tr}[L+R]^2
\label{hls1}
\eea
$a$ is a parameter which is not fixed by the theory and $f_\pi$
is the usual pion decay constant (92.41 MeV).
The covariant derivative is 
\be
D_\mu\xi_{L,R}=\pa_\mu\xi_{L,R} -ig V_\mu\xi_{L,R}+ie\xi_{L,R}A_\mu Q
\label{hls2}
\ee
where $A_\mu$ is the photon field, $V_\mu$ the vector meson field matrix
defined above and $Q= {\rm Diag} (2/3,-1/3,-1/3)$ is the quark charge matrix,
$e$ is the unit electric charge and $g$ is the universal vector
meson coupling \cite{HLS}. Finally, one generally chooses
 the ``unitary'' gauge, for which
\begin{equation} 
\xi_R=\xi_L^{\dagger}=\xi=\exp{({iP}/{f_{\pi}})}.
\label{hls3}
\end{equation}
The standard VMD model is obtained by setting $a=2$ in the
HLS Lagrangian. However, several studies of the pion form
factor \cite{ben4,cmdpi} favor $a \simeq 2.4$, quite inconsistent with 2.
A simultaneous analysis of  light meson radiative decays and 
vector meson leptonic decays \cite{rad,mixing} finds $a \simeq 2.4 -2.5 $, 
quite consistent with pion form factor studies. 

The HLS Lagrangian is given in expanded form in Ref. \cite{heath}
(see Eq. (A1), where the pseudoscalar kinetic energy term
has been omitted). For the purpose of the present paper, it should be noted
that the pseudoscalar singlet field $\eta_0$ undergoes  no interaction
and only occurs in the (omitted) kinetic energy term.

\subsection{SU(3) Breaking Mechanism (FSB) of the HLS Model}
\label{su3brk}

SU(3) symmetry breaking (FSB) of the HLS Lagrangian has been introduced
by Bando, Kugo and Yamawaki \cite{BKY} (already referred to as BKY) and
originates from Refs. \cite{HLS,BKY}.  Brief accounts and some 
new developments can be found in Refs.~\cite{heath,BGP},
connected more precisely with the anomalous sector \cite{FKTUY}.
We refer the reader to Refs.~\cite{rad,conf,BKY,heath,BGP} for 
detailed analyses of the properties of known variants of the BKY 
breaking scheme. Here we will only sketch the so--called
new scheme detailed in Ref. \cite{heath}. Basically, the
BKY breaking of SU(3) symmetry is performed by modifying
Eqs. (\ref{hls1}) in the following way
\be
{\cal L}_{A,V}=-\frac{f_{\pi}^2}{4}{\rm Tr}[(L\mp R)
(1+(\xi_L\epsilon_{A,V}\xi^{\dagger}_R+\xi_R\epsilon_{A,V}\xi^{\dagger}_L)/2)
]^2.
\label{hls4}
\ee
\noindent which has a smooth unbroken limit. The constant matrices
$\epsilon_{A,V}$  are given by 
Diag(0, 0, $c_{A,V}$). Defining the breaking matrices
$X_{A,V}=$ Diag(1, 1, $1+c_{A,V}$), these Lagrangian terms can be written
\be
{\cal L}_{A,V}=-\frac{f_{\pi}^2}{4}{\rm Tr}[(L\mp R)X_{A,V}(L\mp R)X_{A,V}].
\label{hls5}
\ee

The expanded expression of the BKY broken HLS Lagrangian can be found in Ref. 
\cite{heath} (see Eq. (A5) in the Appendix). One should note, among other 
properties  of this breaking mechanism, that the pseudoscalar singlet field 
$\eta_0$ does not undergo interactions with any of  the other fields, as 
in the unbroken limit. It contributes only to the kinetic energy term in 
${\cal L}_A$ 
\be
{\cal L}_A={\rm Tr}[\pa PX_A\pa PX_A] + \cdots
\label{hls6}
\ee
The basic consequence of this BKY breaking mechanism for FSB is thus
to force a renormalization of the (bare) pseudoscalar field matrix $P$,
$P \ra P'$ 
\begin{equation} 
P'=X_A^{1/2} P X_A^{1/2},
\label{brk1}
\end{equation}
in order to restore the kinetic energy term to  canonical form. 
The field transform in Eq.~(\ref{brk1})
has a smooth limit when $X_A \ra {\bf 1}$. Additionally, we have 
\cite{BKY,heath}
\begin{equation} 
z \equiv  1+c_A=\left(\frac{f_K}{f_{\pi}}\right)^2= 
1.495 \pm 0.030 ~~~,
\label{brk2}
\end{equation}

The quantity $z$ was named $\ell_A$ in Ref.~\cite{rad}. 
We shall also use the notation $Z=1/z\simeq 2/3$ 
in the following, for consistency with expressions written in  
Ref.~\cite{rad}.
It should be noted \cite{heath}, that the field renormalization  
(Eqs.~(\ref{brk1}) and (\ref{brk2})) is required in order  to recover 
the charge normalization  condition $F_{K^+}(0)=1$.

The correspondence with the usual ChPT Lagrangian is  easy to establish.
Indeed, concerning PS fields it turns out to consider the $L_5$ term of the ${\cal L}^{(1)}$
Lagrangian together with ${\cal L}^{(0)}$ and the correspondence is $z=1+8L_5m_K^2/f_\pi^2$;
this gives a quite good approximation  \cite{leutwb}
of the expression for $z=[f_K/f_\pi]^2$ and $L_5 \simeq 2.14~10^{-3}$.

Thus, the BKY breaking mechanism outlined above, results in a 
renormalization of the PS field matrix in clear correspondence with ChPT
expectations. It does not result likewise in a renormalization of the vector
field matrix \cite{BKY,heath}. However, the correspondence 
between the results  of Morpurgo \cite{morpurgo} and the so--called
$K^*$  model \cite{rad} could well indicate that a renormalization of the  vector 
fields is also needed \cite{mixing}, but only shows up in the $K^*$  sector. 
In the context of the present paper,
we are actually  independent of any kind of symmetry breaking
in the vector sector. Let us only mention the main result
of Ref. \cite{mixing} which tells that angular departures from
ideal mixing is an appropriate parametrization of the $\omega/\phi$ system
as long as one deals only with on--shell vector resonances.

\subsection{Nonet Symmetry Breaking (NSB) of the HLS Model}

We aim here at providing a reasonable mechanism for NSB in the PS  sector
within an effective Lagrangian. This was  required in 
order to describe successfully the set of observed $VP \gamma$ and 
$P \gamma \gamma$ radiative decays \cite{rad}.

The main problem faced in the phenomenology of radiative decays is the 
generalization of Eq.~(\ref{brk1}) to the case where NSB is also active. 
At leading order, this is solely determined by the influence of NSB on 
the kinetic energy part of an effective Lagrangian. 

If FSB were absent, we already know that 
$P \ra P'_8 + x P'_0$ is the required field renormalization, {\it i.e.}
the renormalization results in a rescaling of the singlet part of the 
$P$ matrix. This means that NSB should contribute specifically
to the kinetic energy term which would become
\be
{\cal L}_A={\rm Tr}[\pa P\pa P] +  c  {\rm Tr}[\pa P_0\pa P_0]\cdots
\label{hls7}
\ee
in the absence of FSB.

Let us now examine how we might incorporate the singlet  contributions 
into the HLS Lagrangian. As is well known, the symmetry of the HLS 
Lagrangian is larger than SU$(N_f)\times$SU$(N_f)$, it is actually
U$(N_f)\times$U$(N_f)$. However, this is unphysical. 
The extra vector U(1) symmetry conserves baryon number
and is thus desirable; moreover, as remarked above, this is supported
by the data. However, the additional axial U(1) symmetry 
is a problem as it would imply either parity doublets
or a ninth light pseudoscalar (for reviews see 
Refs.~\cite{rod,christos,tH} and recently \cite{feldmann}).
Therefore, reducing the symmetry of the HLS Lagrangian is desirable.
Introducing the chiral field $U\equiv\xi^{\dag}_L\xi_R=\exp(i2P/f_\pi)$
\cite{HLS}, one obvious way is through determinant terms \cite{tH},
\be
{\cal L}={\cal L}_{\rm HLS}+ \frac{\mu^2f^2_\pi}{12}
\ln\det U \cdot\ln\det U^{\dag}
+\lambda\frac{f^2_\pi}{12}\ln\det \pa_\mu U
\cdot\ln\det \pa^\mu U^{\dag}
\label{det}
\ee
\noindent where $\mu$ is a parameter with mass dimension
and we have introduced the dimensionless parameter $\lambda$ to
allow for nonet symmetry breaking. Considering  the chiral transformation
$U\ra g^{\dag}_LUg_R$, we see Eq.~(\ref{det}) is now only
invariant under SU$(N_f)\times$SU$(N_f)$ or when $g_L=g_R$ (i.e., $U_V$),
as desired. Rewriting the Lagrangian we have
\be
{\cal L}={\cal L}_{\rm HLS}+ \frac{\mu^2f^2_\pi}{12}
{\rm Tr}\ln U\cdot{\rm Tr}\ln U^{\dag}
+\lambda\frac{f^2_\pi}{12}{\rm Tr}\ln \pa_\mu U
\cdot{\rm Tr}\ln \pa^\mu U^{\dag}.
\label{Z}
\ee
Now recalling Eqs. (\ref{psfield}) and (\ref{u3sym}),  this can be rewritten
\be
{\cal L}={\cal L}_{\rm HLS}+{\cal L}_{\rm HLS}^{\prime}
\equiv{\cal L}_{\rm HLS}+\frac{1}{2}\mu^2\eta_0^2+\frac{1}{2}\lambda
\pa_\mu \eta_0 \pa^\mu \eta_0
\label{nsk1}
\ee
{as $P_0=\eta_0 {\bf 1}/\sqrt{6}$ and Tr[$T^{1-8}]=0$}. Thus, through 
this breaking of the $U_A(1)$ symmetry, the singlet acquires a mass which is 
nonvanishing in the chiral limit and an additional kinetic term. 
As can be clearly seen, this implementation of NSB only modifies the singlet
contribution to the Lagrangian kinetic energy (and mass term) without 
changing the usual HLS  interaction  Lagrangian  (see Eq. (A1) in Ref. 
\cite{heath}).

It is quite interesting at this point to remark  that the NSB parameter
we introduce can be identified with the $\Lambda_1$ coefficient
of the ${\cal L}^{(1)}$ contribution to  the ChPT Lagrangian \cite{leutw,leutwb}, 
as clear from Rel. (13) in \cite{feldmann} who carries practically
the same notations as ours. Therefore, the kinetic energy  of the Lagrangian,
which mostly determines the PS field renormalization, meets all 
expectations from ChPT. 

Having shown how U$_A(1)$ breaking might lead to an additional
Lagrangian term, ${\cal L}^\prime$ as given in Eq.~(\ref{nsk1}),
we now wish to explore the consequences of this. We are interested in
calculating the axial currents. This can be done through an infinitesimal
(axial) variation $\pa_\mu P^a\ra\pa_\mu P^a+f\pa_\mu\epsilon^a$ 
\cite{victor} 
\be
J_\mu^{A,a}\equiv\frac{\pa\cal L}{\pa(\pa_\mu\epsilon^a)}
=f_\pi\frac{\pa\cal L}{\pa(\pa_\mu P^a)}
=2f_\pi{\rm Tr}[T^aX_AT^bX_A]\pa P^b+\lambda f_\pi\delta^{a0}\pa \eta^0.
\label{nsk2}
\ee
We see the octet components are unchanged, while the singlet component
is affected by a factor of $1+\lambda$. 

\section{An Effective Lagrangian Model with FSB and NSB at First Order}
\label{lagnsb}

For the purpose of the present study, we are interested only in 
the PS kinetic energy part of the Lagrangian in Eq. (\ref{nsk1}), 
which needs to be diagonalized in order to get the explicit transform 
$P \ra P'$, from bare to  renormalized fields. Conversely, it is clear that, 
if NSB vanishes, the kinetic energy of the Lagrangian is rendered  canonical
by the transform in Eq. (\ref{brk1}). Here, any reference to what can 
happen in the V sector is totally irrelevant.

\subsection{Diagonalization of the Effective Lagrangian Kinetic Energy}
\label{diagonalization}

The Lagrangian in Eq. (\ref{nsk1}) has a non--canonical kinetic energy,
which is precisely of the form given in Eq. (\ref{hls7}) with $c=\lambda$.
Putting it into a suitable diagonal form is thus required, in order
to define the physical fields in terms of the unphysical (bare)  field
and get the axial currents in terms of  the physical fields.

It is suitable to perform diagonalization in two steps. The first step
is simply to define an intermediate renormalization step by
$P^{\prime \prime}=X_A^{1/2}PX_A^{1/2}$, which puts
the nonet symmetric part of the kinetic energy  term into canonical 
form. Practically,
this means that pion fields are unchanged in this renormalization, while
the kaon fields absorb a $f_K/f_{\pi}$ factor, as if NSB were absent. 
Concerning  isoscalar mesons, these (first step) renormalized fields can be
expressed in terms of the original (bare) fields through

\begin{equation}
\left [
\begin{array}{ll}
\pi_8^{\prime \prime}\\[0.5cm]
\eta_0^{\prime \prime}
\end{array}
\right ]
= z
\left [
\begin{array}{ll}
~~B & ~-A \\[0.5cm]
-A &~~ ~~C
\end{array}
\right ]
\left [
\begin{array}{ll}
\pi_8\\[0.5cm]
\eta_0
\end{array}
\right ]
\label{step1}
\end{equation}
The parameters $A$, $B$ and $C$ 
depend only on the FSB parameter $z$ already defined and they are
\be
A= \frac{\sqrt{2}}{3}\frac{(z-1)}{z}\simeq 0.16~,~~~
  B=\frac{(2z+1)}{3z}\simeq 0.90~,~~~
C=\frac{(z+2)}{3z}\simeq 0.80~,
\label{nsk3}
\end{equation}
where the numerical values correspond to $z\simeq 3/2$. $A$ can be 
considered as the FSB characteristic size. $C$ and $B$ differ at first 
order in this  breaking parameter since $\sqrt{2}(B-C)=A$. After this 
renormalization  the kinetic energy $T$ is still non--canonical.
 Using Eq.~(\ref{step1}), $T$ can be expressed in terms
of the (intermediate, {\it i.e.} double prime) fields by
\begin{equation}
2 {\rm T}=[\partial \pi_8^{\prime \prime}]^2 +
[\partial \eta_0^{\prime \prime}]^2
+ \lambda [A \partial \pi_8^{\prime \prime}+B\partial 
\eta_0^{\prime \prime}]^2.
\label{nsk4}
\end{equation}
It is useful to define the FSB angle $\beta$:
\begin{equation}
\cos{\beta}=
\frac{B}{\sqrt{A^2+B^2}}~,~~~ 
\sin{\beta}= \frac{A}{\sqrt{A^2+B^2}}.
\label{nsk5}
\end{equation}
Diagonalizing Eq.~(\ref{nsk4}) gives the following  renormalized fields
\bea\non
\pi_8^{\prime}&=&\cos{\beta} \pi_8^{\prime \prime}-\
                    \sin{\beta}\eta_0^{\prime \prime}\\
\eta_0^{\prime}&=&\left[\sin{\beta}\pi_8^{\prime \prime}
			+\cos{\beta}\eta_0^{\prime \prime}\right]
			\sqrt{1+\lambda (A^2+B^2)}.
\label{nsk6}
\eea
These field combinations, which directly follow from the 
eigensolutions of the quadratic
form of Eq.~(\ref{nsk4}), have a smooth limit when 
both FSB and NSB tend to zero
($\pi_8^{\prime} \ra \pi_8^{\prime \prime}$ and 
$\eta_0^{\prime} \ra \eta_0^{\prime \prime}$); when FSB
alone tends to zero ($A \ra  0$) the limit is also smooth
($\pi_8^{\prime} \ra \pi_8^{\prime \prime}$ and 
$\eta_0^{\prime} \ra \eta_0^{\prime \prime} \sqrt{1+\lambda}$).
However, the limit is not smooth when only NSB vanishes; indeed,
Eq.~(\ref{nsk6}) shows that the fields remain rotated by an angle $\beta$
which is non--zero if FSB is still active. 
In order to cure this disease, one can choose 
as final renormalized fields linear
combinations of the solutions in Eqs. (\ref{nsk6}), which have the desired
limit properties and conserve the canonical structure 
of $T$ by the diagonalization. Using $v=\sqrt{1+\lambda(A^2+B^2)}-1$,
these combinations are
\bea\non
\pi_8^{\prime}&=&(1+v\sin^2{\beta}) \pi_8^{\prime \prime}
+v \sin{\beta}\cos{\beta}\eta_0^{\prime \prime}\\
\eta_0^{\prime}&=&v \sin{\beta}\cos{\beta}\pi_8^{\prime \prime}
+(1+v\cos^2{\beta}) \eta_0^{\prime \prime}
\label{nsk7}
\eea

\subsection{Field Transformation At First Order}
\label{truncation}

We can use directly the exact transformation given by Eqs. 
(\ref{nsk7}); however, our breaking procedure is actually 
leading order in all breaking parameters. This is illustrated
by having shown the  connection between our breaking  procedure
and the ${\cal L}^{(0)}+{\cal L}^{(1)}$ ChPT Lagrangian.
Therefore, it is meaningful to truncate the field tranform
above at leading order.

Truncating $v$ at first order, 
we have  $v\simeq \lambda (A^2+B^2)/2\simeq \lambda B^2/2$, 
and then
\begin{equation}
\left [
\begin{array}{ll}
\pi_8^{\prime}\\[0.5cm]
\eta_0^{\prime}
\end{array}
\right ]
=
\left [
\begin{array}{ll}
\displaystyle 1 + \frac{\lambda}{2}
A^2&~~~~~ \displaystyle ~~~~~\frac{\lambda}{2}AB\\[0.5cm]
 \displaystyle \frac{\lambda}{2}AB & ~~~~~~
\displaystyle (1+\frac{\lambda}{2}B^2)
\end{array}
\right ]
\left [
\begin{array}{ll}
\pi_8^{\prime \prime}\\[0.5cm]
\eta_0^{\prime \prime}
\end{array}
\right ]
\simeq
\left [
\begin{array}{ll}
\displaystyle 1 & ~~~~~~~0 \\[0.5cm]
 0 & \displaystyle  ~~~~(1+\frac{\lambda}{2}B^2)
\end{array}
\right ]
\left [
\begin{array}{ll}
\pi_8^{\prime \prime}\\[0.5cm]
\eta_0^{\prime \prime}
\end{array}
\right ]
\label{nsk7b}
\end{equation}
The last relation is obtained by removing breaking terms of order greater
than 1. 
Using Eqs.~(\ref{step1}) and (\ref{nsk7b}),
we can approximate the physical field (prime) combinations 
in terms of the bare fields by
\begin{equation}
\left [
\begin{array}{ll}
\pi_8^{\prime}\\[0.5cm]
\eta_0^{\prime}
\end{array}
\right ]
\simeq
\left [
\begin{array}{ll}
\displaystyle B & \displaystyle -A \\[0.5cm]
 \displaystyle -A (1+\frac{\lambda}{2}B^2)& \displaystyle C 
(1+\frac{\lambda}{2}B^2)
\end{array}
\right ]
\left [
\begin{array}{ll}
\pi_8\\[0.5cm]
\eta_0
\end{array}
\right ]
\simeq 
\left [
\begin{array}{ll}
\displaystyle B & \displaystyle -A \\[0.5cm]
 \displaystyle -A (1+\frac{\lambda}{2})& \displaystyle C (1+\frac{\lambda}{2})
\end{array}
\right ]
\left [
\begin{array}{ll}
\pi_8\\[0.5cm]
\eta_0
\end{array}
\right ]
\label{step2}
\end{equation}

This is the physical (first order) approximation
which corresponds to the field renormalization used in 
Ref.~\cite{rad}
and recalled in Eq. (\ref{renorm}). The last matrix expression 
in Eq.~(\ref{step2}) is obtained by remarking
that $\lambda B^2$ differs from  $\lambda$ by terms 
of order $\lambda A$ and then 
is legitimate to neglect them at first order in further computations. 
The $A\lambda$
term in the lower leftmost matrix element is kept for consistency, but clearly
plays a negligible role.

Therefore, the field renormalization on which the study of 
Refs.~\cite{rad,conf,mixing} relies is obtained from a Lagrangian  
model by truncating at first order in the breaking
parameters. This provides an excellent fit to all 
light meson radiative decays,
as can be seen from Ref.~\cite{rad} and as will
be shown below (see Subsection \ref{newfit}). 
We have checked that  a fit to radiative decays  performed
as in Ref. \cite{rad} using  the exact field transformation
in Eq.~(\ref{nsk7}) instead of its first order approximation
in Eq.~(\ref{step2}) gives indeed an improvement in fit quality, but
negligible{\footnote{The improvement is larger if ones replaces
the PDG value \cite{PDG98} for the rate $\phi \ra \eta' \gamma$
by the mean value of all presently available measurements~!
}}. {From} these expressions, 
it is also clear that the NSB  parameter $x$  \cite{rad}   is actually
\begin{equation}
x=1-\frac{\lambda}{2}B^2 \simeq \frac{1}{\sqrt{1+\lambda B^2}} 
\Longrightarrow \lambda \simeq 0.20 - 0.25 ,
\label{xval}
\end{equation}
using the reference value for $x$ (see Eq.~(\ref{fitval})). 
We see the parameter $\lambda$ is small. 
In determining the accuracy
and systematic errors, the neglected orders of magnitude should be
estimated from the values of $A$ [FSB] and $(1-x)$ [NSB]. It should be noted
that, even if $x$ carries prominently the information of NSB,
it is somehow influenced by FSB as $B=1+A/\sqrt{2}$. 

In what follows we shall approximate the change of fields by its 
expression at first order in the breaking parameters, which can be written
\begin{equation}
P=X_A^{-1/2} (P'_8+xP'_0)  X_A^{-1/2}
\label{renorm}
\end{equation} 
The accuracy of this expression relative to the
Lagrangian defined above can be estimated at $\simeq$ 5\% by analyzing the magnitude of the
neglected terms in Eq. (\ref{step2}).  

\section{The VMD Description of $\eta/\eta' \rightarrow \gamma \gamma$ Decays}
\label{anomalous}

Following FKTUY \cite{FKTUY},
the anomalous  U(3) symmetric Lagrangian describing
$PVV$ interactions is
\begin{equation}
{\cal L}=- \frac{3 g^2}{4 \pi^2 f_{\pi}} 
\epsilon^{\mu \nu \rho \sigma} {\rm Tr}[\pa_\mu V_\nu\pa_\rho V_\sigma P].
\label{fktuy}
\end{equation}
The $P V \gamma$ and $P \gamma \gamma$ transitions amplitudes are obtained
from this Lagrangian and  the non--anomalous HLS Lagrangian, needed 
in order to describe the direct transition of vector mesons to photons.
This non--anomalous HLS Lagrangian is  given in its expanded
form in Ref. \cite{heath}. It should only undergo the field renormalization
of Eq. (\ref{renorm}), valid at first order in the (two) breaking parameters.

\begin{figure}[htbp]
     \epsfig{angle=0,figure=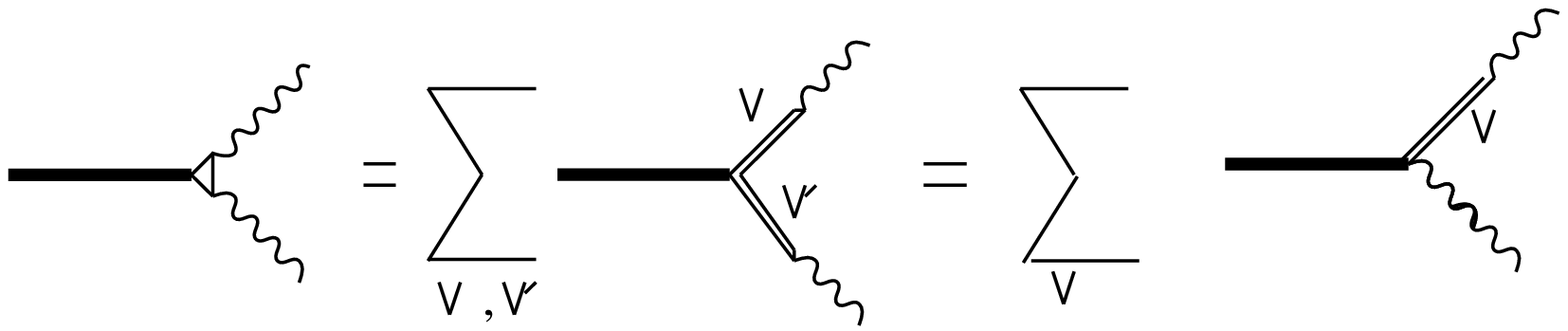,width=0.9\linewidth}
{\caption 
.Graphical representation of the relation among various kind
of coupling constants. $V$ and $V'$ stand for the lowest lying
vector mesons ($\rho^0$, $\omega$, $\phi$); the internal 
vector meson lines are propagators at $s=0$ and are approximated
by the corresponding tabulated \cite{PDG98} masses squared.
\label{graph}}
\end{figure}

The HLS model contains the Vector Meson Dominance
(VMD) assumption (for a review see \cite{review}); 
it thus gives a way to relate the 
radiative decay modes $VP\gamma$ to each other and to the
$P\gamma \gamma$ decays  for light mesons, by giving a precise
meaning to the equations sketched in Fig. \ref{graph}. 

Propagating the field renormalization in Eq. (\ref{renorm}) down to the FKTUY 
Lagrangian of Eq. (\ref{fktuy}) gives
\begin{equation}
{\cal L}=- \frac{3 g^2}{4 \pi^2 f_{\pi}} 
\epsilon^{\mu \nu \rho \sigma} {\rm Tr}[\pa_\mu V_\nu\pa_\rho 
V_\sigma X_A^{-1/2}P'X_A^{-1/2}].
\label{wz2}
\end{equation}
Then, the VVP Lagrangian is changed in a definite way by the renormalization
procedure.

The expanded form of the Lagrangian  in Eq.~(\ref{wz2}) is given in the 
Appendix of Ref.~\cite{rad}. The expressions for the two--photon decay 
widths of the $\eta$ and $\eta'$ mesons can be derived from this and
the non--anomalous Lagrangian. They are\cite{rad}
\bea
\non
G_{\eta \gamma \gamma} &=&  -\displaystyle \frac{\alpha_{em}}{\pi \sqrt{3} 
f_{\pi}}
\left [ \frac{5-2Z}{3}\cos{\theta_P}-\sqrt{2} \frac{5+Z}{3}x \sin{\theta_P} 
\right ],\\
\label{cc4}
G_{\eta' \gamma \gamma} &=&  -\displaystyle \frac{\alpha_{em}}{\pi \sqrt{3} 
f_{\pi}}
\left [ \frac{5-2Z}{3}\sin{\theta_P} + \sqrt{2} \frac{5+Z}{3}x 
\cos{\theta_P} \right ],\\
G_{\pi^0 \gamma \gamma} &=& -\displaystyle  
\frac{\alpha_{em}}{\pi  f_{\pi}},
\non
\eea

\noindent where $Z=1/z=[f_{\pi}/f_K]^2$. Actually, the last expression in Eq.~(\ref{cc4}) 
is a  normalization condition which allows us to fix the numerical coefficient 
in Eq. (\ref{wz2}). It is clear that Eq.~(\ref{cc4}) gives the two--photon decay 
widths in terms of $f_{\pi}$, $f_K$, $x$ and only one mixing angle, $\theta_P$. 
This will be frequently referred to as the wave--function mixing angle 
(see Eq.~(\ref{psrot})). As normal these equations do not (and should not) 
depend on the vector meson parameters ($g$ and $\delta_V$).

Thus, using standard Feynman rules, the HLS model provides definite 
expressions for the two--photon couplings of the pseudoscalar mesons, 
through its anomalous (FKTUY) sector. These expressions exhibit the 
traditional form \cite{DHL,GILMAN,chan} originally obtained through 
Current Algebra. These couplings are related to partial widths by
\begin{equation}
\Gamma(X \rightarrow \gamma \gamma) =
\frac{M_X^3}{64 \pi} |G_{X\gamma \gamma}|^2~~~,~~X=\pi^0 , ~\eta, ~\eta '~~~.
\label{wz5}
\end{equation}

As a  test, one can fit the parameters $x$ and  $\theta_P$ 
solely through radiative decays of the type 
$VP\gamma$ and use these values and their associated
errors to predict the values for the two--photon decay widths of 
the $\eta$ and $\eta'$ mesons. The fit values used for these computations 
\cite{rad} are $x=0.917 \pm 0.017$ and 
$\theta_P=-10.41^{\circ} \pm 1.21^{\circ}$.
The results are given in Table 1 and clearly illustrate that 
the expressions in Eq.~(\ref{cc4}) are valid and that the $VP\gamma$ processes 
accurately {\it predict} the two--photon decay widths. 

Stated otherwise, one does not need more than
one angle  ($\theta_P$) in order to describe the $\eta$  and $\eta'$ 
radiative decays and this receives an especially strong support from all $VP\gamma$ modes.
Additionally, despite claims \cite{DHL,DGH,pich1}, this angle is found to be
$\simeq -10^{\circ}$, in apparent (as will be seen) inconsistency with the ChPT expectation
of  $\simeq -20^{\circ}$ \cite{DHL,DGH,GL85}.

\begin{center} 
\begin{tabular}{|| c  | c  | c | c || c | c ||}
\hline
\hline
\hhhc Mode & VMD Fit  &  PDG & Comment & Global Fit Quality & ($x,\theta_P$)\\
\hhhc ~~& Prediction  &  Value & ~ & $\chi^2/dof$ Probability & Correlation\\
\hline
\hline
 \hhhd~~~& &$0.514\pm0.026$&$\gamma\gamma$&11.07/10(35\%)&$-0.34$\\[0.5cm]
$\eta \rightarrow \gamma \gamma$ [keV]&
$0.464 \pm 0.026$ & $0.46 \pm 0.04$&PDG mean&9.14/10(52\%)&$-0.49$\\[0.5cm]
 ~~~&  & $0.324 \pm 0.046$ & Primakoff & 14.82/10 (13\%)& $-0.55$\\[0.5cm] 
\hline
$\eta' \rightarrow \gamma \gamma$\hhhu[keV] &
$4.407 \pm 0.233$ & $4.27 \pm 0.19$ & PDG mean &  ~ &~\\ 
\hline
\end{tabular}
 
\parbox[t]{16.0cm}{ \hhht
      {\bf Table 1} :  Partial decay widths of the $\eta/\eta'$
      mesons, as reconstructed solely from fits to the radiative decays
      $VP\gamma$ (leftmost data column) and  their direct  measurements
      \cite{PDG98} (second data column). The third data column displays
      fit quality parameters when using the corresponding 
      $\eta$ measurement. The rightmost data column gives the correlation
      coefficient ($x,\theta_P$) in the corresponding case.
}  
\end{center}

A comment is of relevance concerning the data on $\eta \ra \gam$.
One clearly sees that the $VP\gamma$ modes considered altogether
clearly prefer the PDG recommended (mean) value to either of the
homogeneous reported measurements. Therefore, one may guess that
the (single) Primakoff effect measurement and the (fourfold) 
$\gamma \gamma$ measurement, both suffer from systematic errors  
in opposite directions. This guess
is supported by the recent direct measurement of the $\eta \ra \gam$
branching fraction \cite{abegg}  39.21\% $\pm$ 0.3\%,
quite consistent with the PDG mean value. We shall revisit
this issue in Section \ref{feedback}.

In order to substantiate the relative quality of the three data given in
Table 1, we have redone the global fit, as described in Ref.~\cite{rad},
changing only the $\eta \ra \gam$ data. The corresponding
fit information is given in the rightmost pair of data columns in 
Table 1. Even  if the fit probabilities are all quite acceptable, it is clear
that the PDG recommanded value is indeed preferred by the full set of $VP\gamma$
decay modes. For this reason
we use, from now on, the corresponding best fit results as reference values:
\begin{equation}
x=0.902 \pm 0.018~,  ~~~\theta_P=-10.38^{\circ} \pm 0.97^{\circ}
\label{fitval}
\end{equation}

\section{The WZW Description of $\eta/\eta' \rightarrow \gamma \gamma$ Decays}
\label{wzwl1}

Starting from broken HLS and FKTUY, 
the VMD model of Ref.~\cite{rad} recovers the traditional form for the 
two--photon decay amplitudes,
({\it i.e.} the one mixing angle expressions of
Current Algebra \cite{DHL,GILMAN,chan}).
Using these standard expressions, one indeed gets  
through identification with our Eqs.~(\ref{cc4})
\be
\frac{f_{\pi}}{\overline{f_8}}=\frac{5-2Z}{3}~,\,\,\,\,\,\,
\frac{f_{\pi}}{\overline{f_0}}=\frac{5+Z}{6}~x~,
\label{wz6}
\ee
where $Z=[f_{\pi}/f_K]^2$, and  $\overline{f}_{0,8}$
denote the (Current Algebra) singlet and octet decay constants; we 
have already defined $\theta_P$, the (single) mixing angle occurring 
in this  approach. The $\overline{f}_{0,8}$ are named
$\gamma \gamma$ decay  constants in \cite{kekez}.

It  is easy to check that  Eqs. (\ref{cc4}) and (\ref{wz6}) can 
be derived directly from the WZW Lagrangian \cite{WZW1,WZW2}. Indeed, 
this can be written
\begin{equation}
{\cal L}_{WZW} = -\frac{N_c e^2}{4 \pi^2 f_{\pi}} 
\epsilon^{\mu \nu \rho \sigma}
\partial_{\mu}A_{\nu} \partial_{\rho}A_{\sigma} {\rm Tr} [Q^2P]
\label{wzw}
\end{equation}
(with $N_c=3$) where $Q=$ Diag(2/3,--1/3,--1/3)  is the quark charge matrix,
$A$ is the electromagnetic field
and $P$  is the bare pseudoscalar field matrix. Changing to 
the renormalized field $P'$ through Eq.~(\ref{brk1}) allows us to recover  exactly
the couplings in  Eq.~(\ref{cc4}).

This illustrates clearly that, what is named $f_8$
in the Current Algebra \cite{chan} expressions for
$\eta/\eta'$ decays to two photons, can be expressed solely in 
terms of $f_{\pi}$ and $f_K$, in a way which fixes its value to 
$\overline{f_8}=0.82 f_{\pi}$. Correspondingly, we have 
$\overline{f_0}=1.17 f_{\pi}$ which includes a  correction 
of approximately 10\% due to nonet symmetry breaking.
The fact that the WZW Lagrangian leads to the same results as the FKTUY 
Lagrangian simply states their expected equivalence when deriving two--photon 
decay amplitudes. Stated otherwise, the structure of Eqs. (\ref{cc4})
depends only on the BKY breaking $X_A$, with   a small influence
of PS NSB. 

However, the SU(3) sector of Chiral Perturbation Theory (ChPT)
\cite{DHL,GL85,DGH} is well known to predict $f_8/f_{\pi} \simeq 1.25$ 
and a mixing
angle of $\simeq -20^{\circ}$. Then, the question is whether there is an 
inconsistency with respect to ChPT, or if there is a mismatch among 
definitions in ChPT ($f_{0,8}$) and in the VMD/WZW approach
($\overline{f}_{0,8}$), after  symmetry breaking. More precisely,
the question is whether $\overline{f}_{8}$ and $\overline{f}_{0}$
have actually the meaning of decay constants.

Before closing this Section, it is interesting to compare
our expressions for $\overline{f}_{0,8}$ with the corresponding
ones in \cite{kekez} and \cite{scadron3}. Indeed, by identification
of our Eqs. (\ref{wz6}) with Eqs. (26), (31) and (32) of Ref. \cite{kekez}, 
we get

\begin{equation}
\displaystyle \frac{\widetilde{T}_{s \overline{s}}(0,0)}{\widetilde{T}_{\pi^0}(0,0)}= 
\left [ \frac{f_\pi}{f_K} \right ]^2=\frac{2}{3}
\label{KEKEZ}
\end{equation}
for the  ratio of their reduced  amplitudes. Their own numerical estimate 
for this ratio is 0.62, in quite good agreemeent with ours. In order to reach
Eq. (\ref{KEKEZ}), we have stated $x=1$ in our own expressions. So, one can
consider that the present Eq. (\ref{wz6}) extends the results of Ref. \cite{kekez}
(and  \cite{scadron3}) to the case when nonet symmetry is broken; the correction is
however minor in this realm.

\section{A ChPT Description of 
$\eta/\eta' \rightarrow \gamma \gamma$ Decays}
\label{chpt}

We have seen above that, in the
VMD procedure developed in Refs.~\cite{rad,heath,conf},
the expressions for $f_8$, $f_0$ and 
the mixing angle are the same as those obtained
from the matrix elements for 
$\langle\gamma \gamma|{\cal L}_{WZW}|\eta\rangle$
and $\langle\gamma \gamma|{\cal L}_{WZW}|\eta'\rangle$.
We have denoted the parameters, obtained in this manner,
$\overline{f}_{0,8}$  and $\theta_P$. In ChPT, 
however, the corresponding quantities 
($f_{0,8}$, $\theta_8$, $\theta_0$) are  defined
through other matrix elements, namely 
$\langle 0|\partial^{\mu} J_{\mu}^{8,0}| \eta \rangle$
and $\langle0|\partial^{\mu} J_{\mu}^{8,0}|\eta'\rangle$, 
where the $ J^{8,0}$ are the axial currents.
It seems, however, traditionally
admitted \cite{DHL,DGH,holstein} that both sets of definitions 
necessarily coincide.
It is this last property which is addressed now.

\subsection{Usual ChPT Parameters from Broken VMD}

The axial current defined by Eq.~(\ref{nsk2}),
\be
J^{a}_{\mu}=\displaystyle -2f_{\pi} \left\{ {\rm Tr}[T^aX_A \partial_{\mu} 
P X_A]
+\delta^{a0}\lambda \partial_{\mu} P_0] \right \} 
\ee
can be rewritten in terms of the physical fields, through the 
transformation
in Eq.~(\ref{renorm}) (and also Eq.~(\ref{step2}) 
for the isoscalar
sector). We can write the matrix elements 
$\langle0|J^{a}_{\mu}|P^{{\prime} a}\rangle$
for $a=1, \cdots ~7 $ 
and get the corresponding decay constants:
\begin{equation}
\langle0|J^{\pi/K}_{\mu}|\pi/K(q)\rangle=if_{\pi/K}~q_{\mu}
\label{chpt1}
\end{equation}
for pions and kaons, taking into account the expression for $z$.
For the isoscalar sector, we get 
\bea
\non
J^8_{\mu}&=& \frac{1+2z}{3} f_{\pi} \partial_{\mu} \pi^R_8
+\frac{\sqrt{2}}{3}(1-z) f_{\pi} \partial_{\mu} \eta_0^R,\\
J^0_{\mu}&=& \frac{\sqrt{2}}{3}(1-z) f_{\pi} \partial_{\mu} \pi^R_8
+ \frac{2+z}{3} f_{\pi} x(1+\lambda)\partial_{\mu} \eta_0^R,
\label{chpt2}
\eea
with obvious notations, and where we have used 
$z=[f_K/f_{\pi}]^2=1/Z$. {From} above, 
we know that $x=1/\sqrt{1+\lambda B^2}$, is 
influenced by FSB ($B^2 \simeq 0.8$). Moreover, the occurrence 
of simply $(1+\lambda)$
-- without any dependence upon $z$ -- is certainly due to the fact that NSB
in the Lagrangian of Eq. (\ref{nsk1}) does not undergo SU(3) breaking effects. 
Therefore, it is consistent to consider  
that $1+\lambda \simeq 1/x^2$ and
make the (first order) approximation $x(1+\lambda)=1/x$.

One should note the occurrence in Eq.~(\ref{chpt2}) 
of singlet field contributions
to the octet axial current and, conversely, of  octet field contribution
into the singlet axial current. Additionally, these terms 
vanish in the limit of unbroken SU(3) flavor symmetry ($z=1$) as expected.

These  axial currents  allow to define the  following matrix elements
\bea\non
\langle0 J_{\mu}^8|\pi^8(q)\rangle=i f_8 q_{\mu}\,,\,\,\,\,&&
\langle0|J_{\mu}^0|\eta^0(q)\rangle=i f_0 q_{\mu}\\
\langle0|J_{\mu}^8|\eta_0(q)\rangle=i b_8 q_{\mu}\,,\,\,\,\,&&
\langle0|J_{\mu}^0|\pi^8(q)\rangle=i b_0 q_{\mu}
\label{chpt3}
\eea
with 
\bea\non
f_8 =\frac{(1+2z)}{3}f_{\pi}= (1.33 \pm 0.02)f_{\pi\,},\,\, \,\,
&&  b_8 = \frac{\sqrt{2}}{3}(1-z)f_{\pi}=(-0.24\pm 0.01) f_{\pi}\\
f_0= \frac{(2+z)}{3x}f_{\pi}=(1.29 \pm 0.03)f_{\pi}\,,\,\,\,\,
&& b_0 = \frac{\sqrt{2}}{3}(1-z)f_{\pi}=(-0.24\pm 0.01) f_{\pi}
\label{chpt4}
\eea
where the quoted errors are statistical only. So, at leading order in NSB,
we have $b_0=b_8$.

One readily observes a mismatch between the VMD/WZW definition for
$f_8$ and $f_0$ (see Eq.~(\ref{wz6})) and the ChPT definitions 
above; this mismatch is both algebraic and numerical. 
Otherwise, {\it this} $f_8$ corresponds to the standard ChPT definition 
and it has its expected value \cite{DHL,GL85,DGH,holstein}. 

In order to switch to the matrix elements for 
$\langle0|J_{\mu}^{0,8}|\eta/\eta'\rangle$, we use Eq. (\ref{psrot})
together with  the notations of
Kaiser and Leutwyler \cite{leutw,leutwb}
\begin{equation}
\langle0|J_{\mu}^{0,8}|\eta/\eta'(q)\rangle 
=iF^{0,8}_{\eta/\eta'}q_{\mu}
\label{chpt5}
\end{equation}
and find
\bea\non
F^{8}_{\eta}~&=&~~F^{8}\cos{\theta_8}~=~f_8\cos{\theta_P}-b_8\sin{\theta_P}
~=~~(1.269 \pm 0.008) f_{\pi}\\
\non
F^{8}_{\eta'}~&=&~~F^{8}\sin{\theta_8}~=~f_8\sin{\theta_P}+b_8\cos{\theta_P}
~=-(0.472\pm 0.021) f_{\pi}\\
\non
F^{0}_{\eta}~&=&-F^{0}\sin{\theta_0}~=~b_0\cos{\theta_P}-f_0\sin{\theta_P}
~=~~(0.001 \pm 0.023) f_{\pi}\\
F^{0}_{\eta'}~&=&~~F^{0}\cos{\theta_0}~=~b_0\sin{\theta_P}+f_0\cos{\theta_P}
~=~~(1.315 \pm 0.026) f_{\pi}
\label{chpt6}
\eea
using the reference parameter values for $x$ and $\theta_P$ 
of Eq.~(\ref{fitval}). It should be stressed here that $F^{0/8}$
and $\theta_{0/8}$ differ from $f^{0/8}$ and $\theta_P$
only by terms of order $b_0$ and $b_8$. These cannot be neglected
consistently if one keeps terms of order $\sin{\theta_P}$, which
are numerically of the same order. Eqs. (\ref{chpt6}) lead to
\bea\non
F^{8}= (1.36 \pm 0.01)f_{\pi} &&~~~~~
F^{0}= (1.32 \pm 0.03)f_{\pi}\\
\theta_8=-20.40^{\circ} \pm 0.96^{\circ}&&~~~~~
\theta_0=-0.05^{\circ} \pm 0.99^{\circ}
\label{chpt7}
\eea

In the exact SU(3) limit ($z=1$), the expressions in Eq.~(\ref{chpt6}) imply
$\theta_0=\theta_8=\theta_P$. Thus, the difference between them is, indeed, 
an effect of SU(3) flavor symmetry breaking, with only a marginal (numerical) 
influence of the nonet symmetry breaking parameter $x$. 

Now, the values given in Eq.~(\ref{chpt7}) can indeed be compared with ChPT 
expectations, as these  expressions correspond to the 
standard ChPT definition of mixing parameters.
The value for $F^{8}$ compares impressively to the parameter free 
prediction  of Ref. \cite{leutwb}  ($1.34f_{\pi}$ with no quoted error).
The value for $F^{0}$ is harder to estimate theoretically  because of its
scale dependence; however, from the information given in 
Refs.~\cite{leutw,leutwb}, 
$F^{0} \simeq 1.3 f_{\pi}$ seems in an acceptable range. The  value
$\theta_8=-20.40^{\circ} \pm 0.96^{\circ}$  is impressively consistent with 
all reported ChPT  
expectations (for example, $-20^{\circ} \pm 4^{\circ}$ from 
Ref.~\cite{GL85}, $-20.5^{\circ}$ from Ref.~\cite{leutw}).

For $\theta_0$ the situation is unclear because the accuracy of the 
reported  theoretical expectation \cite{leutw} $\theta_0 \simeq -4^{\circ}$ 
is lacking. Then, it is not possible to compare rigorously our result
in Eq. (\ref{chpt7}) with it; we show just below that
the difference with ChPT (if any) is due to non--leading terms
in breaking parameters.

\subsection{Further Comparison of VMD with ChPT}

One can ask about the correspondence of the expressions
for the ChPT parameters coming from our VMD Lagrangian  model
of currents with their usual ChPT expressions in terms of
$f_{\pi}$ and $f_K$.  We show here that all expressions
we can get in our HLS--FKTUY approach and all reported
ChPT expectations to which they can be compared coincide
surely at leading order in breaking parameters.
Let us illustrate with a few examples, mostly 
referred to in Ref. \cite{leutw,feldmann}.

{From} Eqs. (\ref{chpt6}), one can easily derive
\begin{equation}
[F^8]^2=[F^8_{\eta}]^2+[F^8_{\eta'}]^2=\left[ \frac{(1+2z)}{3} \right]^2 f_{\pi}^2+
\frac{2}{9} (1-z)^2 f_{\pi}^2~~.
\label{echpt1}
\end{equation}
Let us define
\begin{equation}
\begin{array}{ll}
z=1+2\varepsilon~,~~~ x=1+\delta~.
\end{array} 
\label{echpt2}
\end{equation}
Neglecting terms of order $\varepsilon^2$, 
we have $\varepsilon=f_K/f_{\pi}-1\simeq 0.22$ 
and $\delta=x-1 \simeq -0.10$
as values for these perturbations to exact SU(3) flavor and nonet symmetries.
Using Eq.~(\ref{echpt2}), Eq.~(\ref{echpt1}) gives
\begin{equation}
[F^8]^2=\left [  1 + \frac{8 \varepsilon}{3}+\frac{8\varepsilon^2}{3} +
{\cal O}(\varepsilon^3) \right] f_{\pi}^2~~~.
\label{echpt3}
\end{equation}
which can be rewritten
\begin{equation}
3[F^8]^2=4f_K^2-f_{\pi}^2 + {\cal O}(\varepsilon^2).
\label{echptf1}
\end{equation}
This gives the same leading term as its ChPT expression 
(see  Eq.~(11) in Ref.~\cite{leutw}). 

Concerning $F^0$, Eqs. (\ref{chpt6}) give at leading order
\begin{equation}
[F^0]^2= \left [ (1+\frac{4\epsilon}{3})(1-2\delta)+{\cal O}(\epsilon^2) \right ] f_{\pi}^2=
\frac{2 f_K^2+f_{\pi}^2}{3}(1-2\delta)+{\cal O}(\epsilon^2)
\label{fchptf1}
\end{equation}
which compares quite well with the  Feldmann expression \cite{feldmann}
\begin{equation}
[F^0]^2=\frac{2 f_K^2+f_{\pi}^2}{3}+f_{\pi}^2 \Lambda_1
\label{fchptf2}
\end{equation}
\noindent from which one derives, neglecting terms of order ${\cal O}(\epsilon \delta)$ 
\begin{equation}
\Lambda_1=-2 \delta =~\frac{1}{x^2} -1 = 0.20 \pm 0.04
\label{fchptf3}
\end{equation}
at the mass scale defined by radiative decays. One should note
that this implies that $\Lambda_1$ is nothing but what was named $\lambda$
as it should since nonet symmetry breaking is done in accordance with the ChPT
${\cal L}_1$ Lagrangian. Eq. (\ref{fchptf3}) may serve to estimate, at the same scale, 
the other OZI violating parameters \cite{leutw,feldmann}  $\Lambda_2$ and $\Lambda_3$
from phenomenology{\footnote{
The symbol $\simeq$ is used because theoretical uncertainties of
the relations among the $\Lambda_i$ are not quoted. 
}}:$\Lambda_3 \simeq -0.03 \pm 0.02$, $\Lambda_2 \simeq 0.31 \pm 0.02$.
These quantities are certainly scale dependent \cite{leutw,leutwb}, however 
this dependence is expected smooth \cite{feldmann}.  

On the other hand, 
Eq.~(\ref{chpt6}) also gives
\begin{equation}
F^0 F^8 \sin{(\theta_8-\theta_0)}=f_8b_0+f_0b_8.
\label{echpt4}
\end{equation}
Using the expressions in Eq. (\ref{chpt4}) and in Eq. (\ref{echpt2}), 
this is
\begin{equation}
3 F^0 F^8 \sin{(\theta_0-\theta_8)}=2\sqrt{2} (f_K^2-f_{\pi}^2) 
\left[1 +\varepsilon -\frac{\delta}{2} \right].
\label{echptf2}
\end{equation}
which coincides at leading order with the corresponding quantity
given in Ref. \cite{leutw} (see Eq.~(13) there). 
One should note, however, that
the leading correction is increased with respect to neglecting deviations
from nonet symmetry, from 22\% to  27\%. On the other hand, one can check 
 that
\begin{equation}
F^0 = 1 +\frac{2}{3}\varepsilon  ~~~, 
 ~~{\rm and} ~~F^8= 1 +\frac{4}{3}\varepsilon
\label{echptf20}
\end{equation}
in units of $f_{\pi}$, which also corresponds to the expectation
that $F^0$ and $F^8$ differ only at first non--leading order \cite{leutw}. 
Of course, leading and non--leading in our expressions refers 
to the small perturbation
parameters of our model, $\varepsilon$ and $\delta$ defined above.

Therefore, at first order in breaking parameters, all expressions
(and numerical values) deduced from our HLS broken Lagrangian meet
all expectations of ChPT at the same order, namely, for  $F^0$,
$F^8$, $\theta_8$ and $\theta_ 0$. Furthermore, using our fit
\cite{rad} to radiative decays, we can provide $\Lambda_1$
with a reliable value.

One should however note that higher order corrections play some role;
indeed, at leading order, $F^0$ and $F^8$ differ by about 15\% 
(see Eq.~(\ref{echptf20}) just above), while the numerical value 
in Eq.~(\ref{chpt7}) gives a difference of only $\simeq 3\%$.

The single divergence  -- if any  --   concerns $\theta_ 0$.
However, this is fully due to  higher order corrections,
the ChPT estimates of which  are still missing.  Indeed,
this can be checked directly on Eq.~(\ref{echpt4}) which gives 
$\theta_0-\theta_8 \simeq -20^{\circ}$, while Eq.~(\ref{echptf2}) gives 
$\theta_0-\theta_8 \simeq -15^{\circ}$, as obtained in 
Refs.~\cite{leutw,leutwb}, {\it i.e.} $-14^{\circ}$ to $-16^{\circ}$. 
With this respect, it should be remarked that Eq. (\ref{echptf2}) 
undergoes  higher order corrections of more than 20\%.

The importance of having an improved theoretical estimate of  $\theta_0$
should be stressed. Indeed,  it should allow to  reject
several modellings (see Table 1 in Ref. \cite{feldmann}), 
as the values proposed for $\theta_0$ range between $0^{\circ}$
and $-30^{\circ}$. Our broken HLS--FKTUY
framework  points out that the condition $\theta_0 \simeq 0$ is 
well fulfilled by all data on radiative decays, as
will be further illustrated in Subsection \ref{newfit}.

\subsection{Relations between $\theta_8$ , $\theta_0$ and $\theta_P$}
\label{relation}

Equations~(\ref{chpt6}) and (\ref{chpt4}) together give
\be
\tan{\theta_8}=\frac{f_8\tan{\theta_P}+b_8}{f_8-b_8\tan{\theta_P}}=
\tan{(\theta_P+\varphi_P)}~,
~~  \tan{\varphi_P}=\sqrt{2}\frac{(1-z)}{(1+2z)} 
\label{echpt6}
\end{equation}
which corresponds to $\varphi \simeq -10.02^{\circ}$,
not influenced by NSB at this order.
On the other hand, the same equations provide also
\be
\tan{\theta_0}=\frac{f_0\tan{\theta_P}-b_0}{f_0+b_0\tan{\theta_P}}=
\tan{(\theta_P-\psi_P)}~,
~~  \tan{\psi_P}=\sqrt{2}\frac{(1-z)}{(2+z)} x
\label{echpt6b}
\end{equation}

Eqs. (\ref{echpt6}) and (\ref{echpt6b}) together give
\be
\theta_8+\theta_0=\displaystyle 2 \theta_P+
\frac{\sqrt{2}}{9}(1-z)^2+\frac{\sqrt{2}}{3}(1-z)(1-x) 
\label{echpt6c}
\end{equation}
up to terms of orders ${\cal O}[(1-z)^3]$ and ${\cal O}[(1-z)^2(1-x)]$,
which represent a few percents only.
This makes explicit the connection between VMD/WZW  and ChPT angles.
Th. Feldmann has recently  obtained a weaker form of the above relations
in  \cite{feldmann} (see its Eq. (84)), in the sense that NSB was neglected.

Let us now consider the condition $\theta_0=0$ as exact, which is clearly
close to real life (see Eq.~(\ref{chpt7})). One can easily check
that this condition can be cast into the form
\begin{equation}
\tan{\varphi_P}=\displaystyle K \tan{\theta_P}~,
\label{echpt7}
\end{equation}
where $K=(2+z)/[(1+2z)x]$ differs from unity by only 3\%, which is likely
well inside our (systematic) model errors. 
Therefore, Eq.~(\ref{echpt6}) can be written
\begin{equation}
\displaystyle \tan{\theta_8}=
\tan{\left[\theta_P+\arctan{(K\tan{\theta_P})}\right]} 
\simeq \tan{2\theta_P} ~~.
\label{echpt8}
\end{equation}
Thus we have
$\theta_P=\theta_8/2$ already with an accuracy of order 1.5\%. A slightly more 
accurate expression is obtained by expanding the relation above
\begin{equation}
\displaystyle \theta_8 \simeq (K+1) ~ \theta_P ~~~,
\label{echpt9}
\end{equation}
taking into account the (observed) 
smallness of $\theta_P$. So, at the level of accuracy permitted
by the data, there is a strict equivalence in using $\theta_8$ 
or $\theta_P$. Furthermore, phenomenology indicates that
higher order corrections should decrease the magnitude
of $\theta_0$.
 
\subsection{Estimate of the Quark Mass Ratio}

Since there is now some reason to trust the reliability
of our numerical results in Eq.~(\ref{chpt6}), we can use
the following equation \cite{leutw}
\begin{equation}
3 \left\{ [F^8_{\eta} M_{\eta}]^2+[F^8_{\eta'} M_{\eta'}]^2\right\}=
4 [f_K M_K]^2 \frac{2S}{(S+1)} - [f_{\pi} M_{\pi}]^2 (2S-1)
\label{echpt5}
\end{equation}
in order to extract the ratio $S$ of the strange quark mass to the 
mean value of the non--strange quark masses ($S=m_s/\hat{m}$). This equation
is actually second degree and thus potentially
admits a spurious solution. 
The $M_i$ terms are the corresponding meson masses. 

Using the mass values for neutral mesons and the information given in 
Ref.~\cite{PDG98}, we get
\be
\frac{m_s}{\hat{m}}=21.23 \pm 2.42 ~~{\rm or}~~ 
\frac{m_s}{\hat{m}}=2.5^{+1.3}_{-0.7}
\label{echptf3}
\end{equation}
The first solution compares well to the expectation  of 
Current Algebra (25.9) and to the estimate (26.6) of Ref.~\cite{leutw}~;
it is also in impressive agreement with the A. Pich estimate \cite{pich1,pich2}
$22.6 \pm 3.3$. The magnitude of the uncertainties (about 10\%)
is dominated by the errors on decay constants; errors due to choosing the
neutral masses for the relevant mesons have not been accounted for. 

\section{VMD, the WZW Lagrangian and ChPT}
\label{comparison}

We have shown in Section \ref{lagnsb}
that the HLS model can be consistently extended in order to include nonet
symmetry breaking along with SU(3) breaking effects. The connection with
the ChPT Lagrangian ${\cal L}^0 + {\cal L}^1$ has been exhibited
and proved successful.
It has been shown reasonable to restrict to first order effects.
The analysis of our HLS--FKTUY model with respect to ChPT, tells us
that the BKY SU(3) breaking mechanism \cite{BKY,heath} is  justified.
The most delicate assumption of the model for radiative decays of Ref. 
\cite{rad} is the PS field renormalization in Eq. (\ref{renorm});
the diagonalization procedure above has shown that  it is indeed
perfectly justified at leading order in breaking parameters.

Instead of radiative decays which  exhibit a huge sensitivity \cite{rad}
to NSB (the parameter $x$), the basic ChPT parameters just considered exhibit 
only a marginal influence of NSB. 

Comparing the two sets of parameters, both derived from within 
a common VMD
framework, the single clear conclusion is that there is a 
mismatch between the VMD/WZW
customary definitions of decay constants and mixing angles and that currently 
stated within ChPT. Before commenting on phenomenological issues,
we first show that this mismatch is a pure effect of SU(3) breaking which
could have been foreseen since the work of Kaiser and 
Leutwyler \cite{leutw,leutwb}. 

\subsection{From ChPT back to VMD/WZW}
\indent \indent The question of whether one can move back from the standard
angles and decay constants of (extended) ChPT to the VMD/WZW framework is,
of course, of special relevance. Indeed, we have already shown that, 
starting from
our VMD/WZW model, the observed mixing angle of $\simeq -10^{\circ}$
(for instance) was quite consistent will all expectations of ChPT,
especially $\theta_8 \simeq -20^{\circ}$.
{The proof of the converse, {\it i.e.} deriving the WZW expressions from
the axial anomaly, can be done on general grounds and is outlined
in the Appendix. We detail here the algebra in order to 
illustrate the connection
between ChPT concepts and the standard VMD parameters,  
and also test the consistency
of having used first order approximation for the field transform.}

The basic idea is  to remark that the divergence of axial 
currents is given by the axial anomaly at $q \ra 0$. In the case of 
two--photon decays, it takes the form
\begin{equation}
\displaystyle \langle0| \partial_{\mu} J^{\mu,a}|\gam\rangle
= N \langle0| {\rm Tr}[ T^a F_{\alpha \beta}\tilde{F}^{\alpha \beta}]|
\gam\rangle~~~, ~~~~ a=0,3,8
\label{victor1}
\end{equation}
where the axial currents are given in Eq.~(\ref{nsk2}),
$F_{\alpha \beta}$ is the photon field strength and 
$\tilde{F}_{\alpha \beta}$ its dual, $T^a$ are
the SU(3) flavor matrices and $N$ is a normalization factor. Saturating
the left--hand side of this expression with the nonet $P$ of the (lightest)
pseudoscalar mesons, we have
\begin{equation}
\displaystyle \langle0| \partial_{\mu} J^{\mu,a}|\gam\rangle
=\sum_P \langle0| \partial_{\mu} J^{\mu,a}|P\rangle \frac{1}{M_P^2} G_{P \gam}
= N ~{\rm Tr}[ T^a Q^2]~~~, ~~~~ a~=~0, ~3, ~8
\label{victor2}
\end{equation}
where we have denoted by $G_{P\gam}$ the decay amplitude of pseudoscalar
mesons to two photons and $Q$ is the quark charge matrix.

As we limit the sum to the lowest pseudoscalar mesons, the single
intermediate state for $\partial J^3$ is $P=\pi^0$ and then
\begin{equation}
\displaystyle
 \langle 0 |\partial J^3|\pi^0\rangle \frac{1}{M_{\pi^0}^2} G_{\pi^0 \gam}=
 N~{\rm Tr}[ T^a Q^2]
\label{victor3}
\end{equation}
With
$\langle 0 | \partial J^3|\pi^0\rangle=f_{\pi} M_{\pi^0}^2$, and because
the last term in Eq.~(\ref{cc4}) gives $G_{\pi^0 \gam}$,  Eq.~(\ref{victor3})
provides the normalization $N=6 \alpha /\pi$.

For $\partial J^0$ and $\partial J^8$, there are two possible
intermediate states (the $\eta$ and $\eta'$ mesons) and Eq. (\ref{victor2})
gives:
\begin{equation}
\begin{array}{llll}
 \displaystyle  F^8_{\eta} G_{\eta \gam}+ F^8_{\eta'}G_{\eta' \gam}
& \displaystyle = \frac{\alpha}{\pi \sqrt{3}},\\[0.5cm]
\displaystyle 
\displaystyle F^0_{\eta} G_{\eta \gam} + F^0_{\eta'}G_{\eta' \gam}
& \displaystyle = \frac{ 2 \sqrt 2 \alpha}{\pi\sqrt{3}},
\end{array}
\label{victor4}
\end{equation}
Inverting these relations gives
\begin{equation}
\begin{array}{llll}
G_{\eta\gam} &=\displaystyle \frac{\alpha_{em}}{\pi \sqrt{3}} 
\frac{1}{F^8_{\eta}F^0_{\eta'}-F^0_{\eta}F^8_{\eta'}} 
\left[F^0_{\eta'}-2\sqrt{2}F^8_{\eta'}\right] \\[0.5cm]
G_{\eta'\gam} &=\displaystyle \frac{\alpha_{em}}{\pi \sqrt{3}} 
\frac{1}{F^8_{\eta}F^0_{\eta'}-F^0_{\eta}F^8_{\eta'}} 
\left[-F^0_{\eta}+2\sqrt{2}F^8_{\eta}\right] 
\end{array}
\label{victor5}
\end{equation}

These expressions are the $\eta/\eta' \ra \gam$ amplitudes in terms of the 
Kaiser--Leutwyler parameters $F^{0/8}$ and $\theta_{0/8}$. 
They can be reexpressed
in terms of $\theta_P$  and $f_0$, $f_8$, $b_0$, $b_8$ 
by means of Eq.~(\ref{chpt6})
and become  
\begin{equation}
\begin{array}{llll}
G_{\eta\gam} &=\displaystyle \frac{\alpha_{em}}{\pi \sqrt{3}} 
\frac{1}{f_0 f_8-b_0 b_8} \left[
(b_0-2\sqrt{2}f_8)\sin{\theta_P} + (f_0 -2\sqrt{2}b_8)\cos{\theta_P}
\right],\\[0.5cm]
G_{\eta'\gam} &=\displaystyle \frac{\alpha_{em}}{\pi \sqrt{3}} 
\frac{1}{f_0 f_8-b_0 b_8} \left[
-(b_0-2\sqrt{2}f_8)\cos{\theta_P} + (f_0 -2\sqrt{2}b_8)\sin{\theta_P}
\right].
\end{array}
\label{victor6}
\end{equation}

 These expressions exhibit the standard Current Algebra structure. Having
denoted the coefficients there by $\overline{f}_{0,8}$, we clearly have 

\be
\begin{array}{ll}
\displaystyle
\frac{1}{\overline{f_8}}=\frac{f_0-2\sqrt{2}b_8}{f_0f_8-b_0b_8}~~~, ~~~
&\displaystyle \frac{1}{\overline{f_0}}=-\frac{1}{2\sqrt{2}} \frac{b_0-2\sqrt{2}f_8}{f_0f_8-b_0b_8}
\end{array}
\label{ca1}
\ee 

This proves that $\overline{f_8}$ and $\overline{f_0}$ would coincide with the 
$\eta/\eta'$ decay constants $f_8$ and $f_0$ , only if $b_0$ and $b_8$ were zero,
{\it i.e.} if SU(3) were not broken~; in this case we would have $f_8=xf_0=f_\pi$.
Using the expressions in Eq.~(\ref{chpt4}) and truncating at leading order
in $(z-1)$ and $(x-1)$, it is easy to check that we get
the $\overline{f_i}$ as given in Eq.~(\ref{wz6}).

The expressions in Eq.~(\ref{ca1}) are interesting in this regard:
they clearly show that the mismatch originates from the fact that
$b_0$ and $b_8$ are non--zero when SU(3) symmetry is broken ($z \ne 1$),
which is basically the point of Refs.~\cite{leutw,leutwb}. Therefore,
the usual expressions of Current Algebra do not  directly give
the isoscalar meson decay constants when SU(3) is broken. 

Nevertheless, this does not prevent Current Algebra from being the most 
economic formulation for the study of radiative decays, as it
involves only two parameters ($\theta_P$ and $x$) instead of four
highly correlated parameters (see Eqs. (\ref{victor5})). Moreover,
we shall see in the next Section how  $x$ and $\theta_P$ are actually
related by the observed smallness of $\theta_0$. This means that the Current Algebra 
formulation, suitably used, depends (practically) on a single parameter;
this can be chosen equivalently as either of $x$ (or $\Lambda_1$) or $\theta_P$.
In this case,  the two--photon decay amplitudes for $\eta$ and $\eta'$
become a constrained system  (2 equations, 1 parameter)!

If one takes into account the smallness of $\theta_0$ expected from ChPT 
and from VMD estimates (see Eq.~(\ref{chpt6})), Eq.~(\ref{victor5})
provides quite an  interesting result
\begin{equation}
\begin{array}{llll}
G_{\eta\gam} + G_{\eta'\gam} \tan{\theta_8} = \displaystyle 
\frac{\alpha}{\pi\sqrt{3}} \frac{1}{F^8 \cos{\theta_8}}
+{\cal O}(\theta_0), \\[0.5cm]
G_{\eta'\gam} = \displaystyle \frac{\alpha}{\pi\sqrt{3}} 
\frac{2\sqrt{2}}{F^0} +{\cal O}(\theta_0) ,
\end{array}
\label{victorZ}
\end{equation}
together with $\theta_P=\theta_8/2+{\cal O}(\theta_0)$.

\subsection{Phenomenological Consequences}

The consequences of the mismatch mentioned above are rather practical.
Usually, the experimental determination
of the pseudoscalar parameters ($f_0$, $f_8$, $\theta$) is derived using
the (quite standard) VMD/WZW Eqs.~(\ref{cc4}), which are nothing but
the former  equations of Current Algebra \cite{DHL,GILMAN,chan}. 

However, generally, the model reconstruction
quality \cite{ben2,holstein} is defined 
by comparing fit results for $\overline{f_8}$,
$\overline{f_0}$ or $\theta_P$ with ChPT numerical expectations ($F^8$,
$F^0$, $\theta_8$). This leads to highly confusing situations
as shown by the discussions in Refs. \cite{kekez,signal} when 
justifying the value found for the mixing angle ($\theta_P \simeq
-13^\circ$)  which mainly differs from our
because of their neglecting NSB in the PS  sector.
The problem is strictly the same with decay constants \cite{kekez}.

Sometimes, numerical ChPT  expectations for
$F^8$ are attributed to what has been named $\overline{f_8}$
in order to  to constrain the WZW two--photon equations 
\cite{DHL,GILMAN,holstein}; this mechanically (and artificially) pushes
the mixing angle to $\simeq -20^{\circ}$, as can be seen
from Fig. 1 in \cite{ball}.

What has been illustrated above is that such a phenomenological approach
and such a theoretical treatment, are  intrinsically inconsistent,
as also noted by   Ref. \cite{feldmann}. 
Stated otherwise, it is meaningless to compare (or constrain)
the  Current Algebra equations (given also by Eqs. (\ref{cc4}))
using the PS decay constants of  ChPT or the  value of the ChPT angle
$\theta_8$, as traditionally done.

To be more specific, we have proved that 
$\overline{f_8}=(0.82 \pm 0.01)f_{\pi}$, 
$\overline{f_0}=(1.17 \pm 0.02)f_{\pi}$
and a (single) $\theta_P \simeq -10^{\circ}$ which describes the $\eta/\eta'$ 
mixing at the wave function level are consistently derived
from VMD and/or the WZW Lagrangian after applying FSB (and NSB).
Moreover, all this is perfectly consistent with
all ChPT expectations at first order in breaking parameters: mainly 
$F^8 \simeq (1.25 - 1.35) f_{\pi}$
and $\theta_8 \simeq -20^{\circ}$. VMD has been able to provide new information
($\theta_0 \simeq 0$ and $F^0 \simeq 1.3 f_{\pi}$, $\Lambda_1 \simeq 0.20$) of 
relevance for ChPT.
A more refined comparison should wait until higher order ChPT estimates
become available.

As a side remark, one should also recall, from Ref. \cite{heath}, that 
SU(3) breaking does not affect the box anomalies for 
$\gamma \pi^+ \pi^- \eta/\eta'$. It is clear, however,
from Eq. (39) in this reference, that nonet symmetry breaking can play some 
(numerically) minor role. Thus, all existing
analyses \cite{ben2,holstein} of the anomaly equations \cite{chan}
have to be redone from scratch, at least for consistency, knowing
that the expected parameter values are not (directly) the ChPT ones.
Related to this point, it is clear now that the Chanowitz equations 
\cite{chan}, correctly understood, do not point any longer to a failure
of QCD, as incorrectly deduced in Ref. \cite{ben2} because of the confusing angle 
problem mentioned above.

\section{Feedback from the ChPT Parametrization}
\label{feedback}

The ChPT parameter values we have obtained
(see Eq.~(\ref{chpt7})) allow for several remarks of
importance which are to be discussed in this Section.

\subsection{A Hidden Relation between $x$ and $\theta_P/\theta_8$}
\label{xtheta}

At the level of accuracy permitted by the whole set
of radiative decays of light mesons, the results gathered in 
Eq.~(\ref{chpt7})
indicate that $\theta_0=0$ is well fulfilled experimentally.
At its level of accuracy ($\theta_0=-0.05^{\circ} \pm 1^{\circ}$),
one can even ask oneself whether this relation is only approximate;
this means that $\theta_0$ does not undergo significant
effects of SU(3) breaking, as opposed to $\theta_8$ and $\theta_P$. 
As remarked in Ref. \cite{leutw}, this also means that, in the sense that
$|\eta\rangle$ is orthogonal to $J^0|0\rangle$, 
the $\eta$ meson is  practically pure octet.
But as shown above, 
the same  $|\eta\rangle$ 
happens also to be 
a mixture of $|\pi^8\rangle$
and $|\eta^0\rangle$  with an angle $\theta_P \simeq -10^{\circ}$. This 
illustrates the duality of definitions from another point of view. 

The numerical result $\theta_0=0$ indicates that the state mixing angle 
$\theta_P$  fulfills
\begin{equation}
\tan{\theta_P} = \displaystyle \sqrt{2} \frac{(1-z)}{(2+z)} ~x
\label{chpt8}
\end{equation}
to good accuracy; this calls for several important remarks.

\begin{itemize}
\item
By providing a definite value for $x$, Eq.~(\ref{chpt8}) allows 
to address the issue of a possible glue content inside the $\eta'$ (see, for 
instance, Table II in Ref. \cite{rad} or Fig. 2 in Ref. \cite{conf}).
Indeed, this equation is purely a consequence of $\eta$ physics
and Refs. \cite{rad,conf} have shown that no glue was required
inside the $\eta$ meson. As a matter of consequence, Eq.~(\ref{chpt8})
is not influenced by a possible glue content in the $\eta'$  meson.

\item
Eq.~(\ref{chpt8}) 
reveals an unexpected {\it algebraic} relation{\footnote{
As a matter of consequences, there is a numerical
correlation ($x,\theta_P$) in fit procedures, which has
been completely missed in our study in Ref.~\cite{rad}.
The correlation coefficient is given in Table 1. This
is smaller than what could have been expected, but
the sign is consistent, if one remarks that $\theta_P$ and $x$ 
carry opposite signs.}} 
between the mixing 
angles $\theta_P/\theta_8$ and the nonet symmetry breaking parameter
$x$. It will be checked explicitly in Subsection \ref{newfit}
using  the whole set of radiative decays. 

\item Eq.~(\ref{chpt8}) allows, for the first time, for
a constrained fit to solely the $\eta \ra \gamma \gamma$
and $\eta' \ra \gamma \gamma$ partial widths. Indeed,  
symmetry breaking effects in terms of singlet and octet components are completely 
determined by the BKY procedure, as shown in Section \ref{wzwl1}. Only $x$
is still free.  Now, Eq. ~(\ref{chpt8}) tells that $\theta_P$ is not an independent 
additional parameter and is fixed by the value $x$. So, the single undetermined parameter 
is either $\theta_P$ or $x$.

We have performed the exercise and got $\theta_P=-10.34^\circ \pm 0.22^\circ$
(corresponding to $x=0.903 \pm 0.017$); using the PDG value of the $\eta$ rate
-- confirmed by the recent measurement of Ref. \cite{abegg}. We get 
$\chi^2/dof= 0.8~10^{-4}/1$ (a 99\% probability), while the Primakov
measurement gives $\chi^2/dof=6.99/1$ (0.8\% probability), and the
$\gamma \gamma$ measurement gives $\chi^2/dof=3.89/1$  (4.8\% probability).
So, the correlation is indeed observed; the reconstruction 
quality indicates however that the Primakoff measurement is
affected by large systematics; this effect is present to a lesser
extend in the measurement of the $\gamma \gamma$ Experiments.
Therefore, a new accurate measurement of $\eta \ra \gamma \gamma$
would be welcome.
\end{itemize} 

To our knowledge,  it is the first time such a relation as Eq.~(\ref{chpt8})
is reported. If $z=1$ (then all $\theta$'s vanish),  $x$ does not vanish but
becomes unconstrained. Traditionally, the wave--function
mixing angle is expressed in terms of PS meson masses~;
our expression (\ref{chpt8}) tells that the mixing angle can also 
be expressed as a function of $f_K/f_\pi$ with some influence
of NSB. 

{From} the point of view of ChPT, it should be stressed that  
$\theta_P$  is the variable  which emphasizes the most
the existence of nonet symmetry breaking; using the terminology
of ChPT, it is directly proportional  to $x=1/\sqrt{1+\Lambda_1}$,
which exhibits the  scale dependence of $\theta_P$.

\subsection{A New Global Fit to Radiative Decays}
\label{newfit}

{From} Eq. (\ref{chpt8}), we have been led to conclude that
$x$ and $\theta_P$ are algebraically related, at least
to a very good approximation. In order to check this in its  full realm,
we have redone the fits given in Ref. \cite{rad}, requiring additionally this
functional relation. This turns out to describe all radiative decays in terms
of only 4 independent parameters ($g$, $\theta_P$,  $\theta_V$, $\ell_T$)
or ($g$, $x$,  $\theta_V$, $\ell_T$),
which is, by far, the most constrained fit of the 14 radiative
decay modes ever attempted{\footnote{As, now, we know
that $\theta_P$ and $\theta_8$ are functionally related, and 
that $\theta_8 \simeq -20^{\circ}$ is equivalent
to  the  favored \cite{rad} $\theta_P \simeq -10^{\circ}$, 
we could, in principle, fix $\theta_8$ (and then $\theta_P$) to its
ChPT expectation.  We have  not perform this exercise, as the accuracy on 
the ChPT estimate of $\theta_8$ is still poor \cite{GL85} and
its sensitivity to NSB somewhat unclear. Nevertheless, it indicates
that, from first principles, one can perform a 
fit to the 14 radiative decays, with remaining free fitting
parameters referring only to vector mesons properties. }}.
We use the constant approximation for $\delta_V$ 
following the conclusion of our study \cite{mixing}.

The fit quality obtained when setting up the constraint 
is $\chi^2/dof=9.14/11$, and does not exhibit
any degradation compared to the fit quality reached when releasing this constraint
$\chi^2/dof=9.13/10$. The difference, in this last case, with Ref. \cite{rad}
is simply  the use inside the fit of the PDG mean value for $\eta \ra \gam$
instead of its mean value from all  experiments (including the Primakoff effect measurement).
In all cases, we have used  the so--called \cite{rad} $K^*$ model,
also commented on in Ref. \cite{mixing}. Therefore, Eq. (\ref{chpt8})
is indeed intrinsically present in the full data set examined sofar.

Practically, the fit returns all parameters at the values obtained
when leaving $x$ and $\theta_P$  unrelated, even
$\theta_P$ which changes from $\theta_P=-10.38^{\circ}\pm 0.97^{\circ}$
to $\theta_P=-10.32^{\circ}\pm 0.20^{\circ}$. The sharp
reduction of the statistical error is an effect of removing
the correlations  by accounting explicitly for the functional
relation in  Eq.~(\ref{chpt8}). The corresponding value for $x$
is 0.901 (to be compared with the fit value $x=0.902$ mentioned
above). This numerical value has an important consequence, as will be 
seen in the next subsection. The fit branching fractions
with $x$ and $\theta_P$ left free or related are given in Table 2
altogether with the data recommended by the PDG \cite{PDG98}, all 
used in the fit procedure.

When setting the $x-\theta_P$ relation, 
we do not observe any degradation in the 
quality of the description of the various branching 
fractions.  Comparing the two sets of predictions in Table 2,
one should note the sharp
reduction of the statistical errors produced by having
switched on the $x-\theta_{P}$ relation in all decay modes
involving the $\eta$ meson. This also affects the modes
involving $\eta'$, though to a lesser extent.

The increased accuracy we observe is not an artifact, but
a trivial consequence of  using independent (fit) variables
instead of correlated ones. This explains why, even if it were
theoretically better motivated, the use of correlated quantities
like $F^0_{\eta/ \eta'}$ and $F^8_{\eta /\eta'}$ is
not recommended in numerical analyses, as this results in unaccurate
uncertainties. 

One should also note that recent measurements \cite{purlatz}
of $\phi \rightarrow \eta' \gamma$ by the  CMD--2 Collaboration at VEPP--2M,
using new $\eta'$ decay modes,
confirm the central value of Ref. \cite{phietp2} rather
than that of Ref. \cite{phietp1} reported by the PDG \cite{PDG98}; the
agreement with the fit values we always get for this mode \cite{rad} 
is thus improved. Indeed, the new measurement 
([$5.8 \pm 1.8]\times 10^{-4}$) reported
by Ref.~\cite{purlatz} (or by Ref.~\cite{phietp2})
would provide a minimum $\chi^2$ smaller than reported above by one unit;
it has not been used in the fit in order to keep consistency with 
all 1998 recommended values \cite{PDG98}.

\subsection{The $x$--$\theta_P$ Relation Kills Glue in the $\eta'$ Meson}

In the previous study of Refs.~\cite{rad,conf}
it was shown that the correlation between glue  and nonet
symmetry breaking was huge. As stated 
several times above, accounting generally 
for glue coupling to the $\eta/\eta'$ system implies that two angles
have to be introduced in addition to $\theta_P$. One ($\beta$) is such that 
$\beta=0$ implies that the $\eta$ meson does not couple to glue, the other
($\gamma$) is such that $\gamma=0$ implies a decoupling of the $\eta'$ from glue.

However Table IV in Ref. \cite{rad} or Fig. 2 in Ref. \cite{conf}, clearly 
show that: {\bf i/} whatever 
the value of the nonet symmetry breaking parameter
$x$, the angle $\beta$ is not observed to deviate sensitively from zero;
{\bf ii/} additionally, for $x \simeq 0.9$  the angle $\gamma$  is also
consistent with zero. 

We have  seen above that it was indeed appropriate to parametrize
PS NSB by $x$ and given its most probable (fit) value
0.901. So, we can conclude that within the picture presented in this paper,
there is no signal for a glue component inside the $\eta$  and the $\eta'$
mesons, but instead there is a significant signal for deviation
from exact nonet symmetry: $x=0.901$ relative to 1 reveals a
$5 \sigma$ significance level.

This is confirmed by performing the fit with $\gamma$ and $x$,
now (numerically) decorrelated because of the functional
relation in Eq.~(\ref{chpt8}). In this case, the fit quality  is strictly 
unchanged  $\chi^2/$dof$=9.14/10$ and the minimum is reached for
$\gamma=-0.02^{\circ}\pm 18^{\circ}$; this 
shows that no glue component inside the $\eta'$ is required
by the data.

The conclusion would be quite different \cite{rad,conf,basu1,basu2,kou} 
if NSB could have been neglected. Indeed the level of correlation between
standard  nonet symmetry breaking (the parameter $x$) and glue is
such that  one can easily misidentify the former as being the later. 
So, even if  our data set cannot exclude the existence of glue
coupled to the $\eta/\eta'$  system, it exclude presently its need. 

\subsection{$\theta_P$ and the Isoscalar Mass Matrix}

In light of the above,
the ChPT picture happens to be consistent
with the standard one angle state mixing scheme of  fields 
(or wave--functions):

\begin{equation} 
\left[
     \begin{array}{l}
     \displaystyle \eta   \\[0.5cm]
     \displaystyle \eta'  
     \end{array}
\right]
=
\left[
     \begin{array}{lll}
\displaystyle \cos{\theta_P} & -\displaystyle \sin{\theta_P} \\[0.5cm]
\displaystyle \sin{\theta_P} & ~~\displaystyle \cos{\theta_P}
     \end{array}
\right]
\left[
     \begin{array}{ll}
     \pi_8\\[0.5cm]
     \eta_0
     \end{array}
\right]
\simeq
\left[
     \begin{array}{lll}
\displaystyle \cos{\frac{\theta_8}{2}} &~~~ -\displaystyle \sin{\frac{\theta_8}{2}} \\[0.5cm]
\displaystyle \sin{\frac{\theta_8}{2}} &~~~ ~~~\displaystyle \cos{\frac{\theta_8}{2}}
     \end{array}
\right]
\left[
     \begin{array}{ll}
     \pi_8\\[0.5cm]
     \eta_0
     \end{array}
\right]
\label{mix1}
\end{equation} 

\noindent 
The most accurate value for $\theta_P$  comes
out from fit to all radiative decays with the $x- \theta_P$
correlation set up. Indeed, from the subsection just above,
we know that it is legitimate to neglect coupling to glue.
Actually,  one cannot assert that glue (or any other singlet
state) is not present inside the $\eta/\eta'$ system, but
what is shown in Refs. \cite{rad,conf} for $\eta$, and
just above for $\eta'$, is that no glue contribution is
required. A kind of minimum complexity argument, then leads
to state $\beta=\gamma=0$ and to decouple glue
from the $\eta/\eta'$ system. The angle value is 
$\theta_P=-10.32^{\circ} \pm 0.20^{\circ}$, a hardly constrained
value. 

With this, it is possible to revisit the determination
of the isoscalar mass matrix \cite{DHL,DGH,holstein}.
Indeed we know that the mass matrix ${\cal M}$:

\begin{equation} 
{\cal M}~~
=
\left[
     \begin{array}{lll}
\displaystyle m_{88}^2 & ~~\displaystyle  m_{08}^2\\[0.5cm]
\displaystyle m_{08}^2 & ~~\displaystyle m_{00}^2
     \end{array}
\right]
\label{mix2}
\end{equation}

\noindent admits the following eigenvectors:

\begin{equation} 
\left \{
\begin{array}{ll}
v_{\eta}=& (\cos{\theta_P}~,-\sin{\theta_P}) \\[0.5cm]
v_{\eta'}=&(\sin{\theta_P}~,~~\cos{\theta_P})
\end{array}
\right. 
\label{mix3}
\end{equation} 

\noindent with eigenvalues $M_{\eta}^2$ and $M_{\eta'}^2$.
This gives $m_{88}^2=0.320 \pm 0.001$, $m_{00}^2=0.898 \pm 0.001$
and $m_{08}^2=-0.109 \pm 0.004$ in units of GeV$^2$.
This corresponds to $m_{00}=0.948  \pm 0.001$ GeV,
and $m_{88}=0.566 \pm 0.001$ GeV, the former  
close to the $\eta'$ mass and the latter  
close to the $\eta$ mass as one might expect from the value
of $\theta_P$. The off--diagonal term can be written $m_{08}^2=-0.45 M_K^2$. 
The solution favored by VMD phenomenology is a very small
deviation from the classical  Gell--Mann--Okubo 
formula\cite{DHL,DGH,holstein}.
More precisely, from:  

\begin{equation} 
M_{\pi^8}^2=\displaystyle \frac{(4 M_K^2-M_{\pi}^2)}{3} \left[ 1+\Delta 
\right]
\label{mix4}
\end{equation} 

\noindent (with $M_{\pi^8}\equiv m_{88}$)  one extracts 
$\Delta \simeq 0.01$, where most of the error is due to
choosing the mass values for $K$ and $\pi$.

\subsection{Broken VMD, ChPT and the Third Mixing Angle}

We have shown that the HLS model, after applying FSB and NSB,
yields the structure of the ChPT Lagrangian ${\cal L}^{(0)}+{\cal L}^{(1)}$.
To be more specific, this concerns the PS kinetic energy part and
the $\Lambda_2$ and $L_8$ terms  \cite{leutwb,feldmann} are not considered;
they would influence the PS mass term which is outside the realm of this
paper. Therefore, one can state that ${\cal L}^{(0)}+{\cal L}^{(1)}$
is equivalent to the HLS Lagrangian broken as shown in Section
\ref{lagnsb} and the relationship exhibited in Section 
\ref{relation} between $\theta_8$ and $\theta_P$ is not accidental.

Expressing the $\eta/\eta' \ra \gamma \gamma$ coupling constants
in terms of $\theta_P$, we have seen that this is the mixing angle 
which traditionally parametrizes the Current Algebra expressions. 
It is clearly a natural parametrization of VMD instead of
the two mixing angles $\theta_0$ and $\theta_8$. For this purpose, it
is quite interesting to compare our Eqs. (\ref{cc4}) with
the corresponding Eqs.(39) in Ref. \cite{feldmann}, expressed
in terms of $\theta_0$ and $\theta_8$.

Therefore, as it  is possible to provide $\theta_P$  a
clear physical meaning through a mathematically well defined procedure, 
it is a quite legitimate choice. Its main virtue is to allow
for a straigthforward handling of symmetry breaking
effects in the VMD Lagrangians (HLS and FKTUY) as shown
in Refs. \cite{heath,rad}. The present paper has shown
that the corresponding parametrization of VMD is easy
to connect with the ChPT conceptual framework.

\subsection{Nonet Symmetry Breaking and the Third Mixing Angle}

Let  us make a  final (practical) remark on NSB. Ref. \cite{rad}, and Ref. 
\cite{ben2} before, clearly proved  that nonet symmetry breaking plays numerically 
a major role in accounting for radiative decays of light mesons within a relatively
simple and constrained framework. Even if small in absolute magnitude
($\delta \simeq -0.10$),  the effect is statistically significant ($\simeq 5\sigma$).
At the present level of experimental accuracy, the data description
quality is sensitive to this improvement. 

Quite interestingly, neglecting NSB in coupling expressions does not 
result in a dramatic change of the fitted mixing angle value, but mostly of
the fit probabilities. Ref. \cite{rad} has thus obtained 
$\theta_P=-14^\circ \pm 1^\circ$ with, however, a degraded fit quality 
($\chi^2/dof = 32/9$). Table 1 in Ref. \cite{rad} clearly
show that all other physics parameters have unchanged fit values. 

This kind of angle value has been obtained in several other approaches~;
their common feature is their neglecting NSB in the pseudoscalar sector. Ref.
\cite{scadron1b} thus obtains $\theta_P=-18.2^\circ \pm 1.4^\circ$
in their fit to $VP \gamma$ processes and $\theta_P=-12.3^\circ \pm 2.0^\circ$
in their fit to the $P \gamma \gamma$ modes{\footnote{One should note
that, even if both fit qualities are {\it separately} good, the two angle values 
are different enough that one can guess that the fit quality of a {\it global} 
description would be certainly degraded and providing $\theta_P \simeq -15^\circ$.
}}. Refs. \cite{scadron2,kekez} within the bound state
approach and dealing with 2--photon processes only also get
a mixing angle of $\theta_P \simeq -12^\circ$. The model of Ref. \cite{gedalin}
finds a fit solution  $\theta_P \simeq -15^\circ$ (with no quoted fit quality)~;
its freedom is however much larger than ours and the relation between their
breaking scheme and ours -- which now is motivated by ChPT concepts -- is unclear.  

Instead, in our approach, the best solutions to separately $VP \gamma$
and $P \gamma \gamma$ processes coincide and both correspond to 
$\theta_P = -10.3^\circ \pm 0.2^\circ$. Additionally, we should
remind that, in our approach, all $P \gamma \gamma$ couplings 
depend at leading order only on $f_K/f_\pi$ and $\theta_P$ {\it or} the 
NSB parameter $x$.
From this point of view, we consider that the result reported
in \cite{ukqcd} (a prefered $\theta_P \simeq -10.2^\circ$) gives
our approach a strong support from lattice QCD estimates.  
 
It should thus be stressed that NSB in the pseudoscalar sector
is exhibited, not so much by sharp changes in  numerical values of
physics parameters, but rather by their improved fit probabilities and, therefore, their
actual accuracy. Indeed, as clear from this Section, and  the previous
ones, it was shown that the main ChPT parameters have values
which are not sharply sensitive to having $x \ne 1$. 

This is also the reason why the numerical results of Ref.  \cite{kekez}
are so close of ours, even if these authors neglect NSB within
the limited data set they consider. To be more specific,
effects of the small value of $\delta$ (or of $\Lambda_1$) are competing with
SU(3) breaking, always by modifying non--leading corrections;
for instance, the magnitude of the correction terms to $\theta_0-\theta_8$
(see Eq.~(\ref{echptf2}), for instance).

\section{Conclusion}
\label{conclud}

In a previous work, we were faced  with
a paradoxical problem.
Using the HLS model and its anomalous FKTUY sector, together
with a definite breaking scheme, it is possible to achieve
quite a satisfactory description of all radiative decays,
including $\eta/\eta' \rightarrow \gamma \gamma$. Within this
framework, it was moreover possible to predict accurately these last
rates, using only numerical information obtained by fitting
the $VP\gamma$ processes in isolation. 
This quite satisfactory pattern was obscured by some strange
results: the (single) pseudoscalar mixing angle was found at 
$\theta_P \simeq -10^{\circ}$ and the octet decay constant 
was $f_8=0.82 f_{\pi}$, both in obvious disagreement with
ChPT expectations. This meets independent analyses
of other authors like Kekez, Klabu\v{c}ar, Scadron, Bramon~; it
received  recently a clear support of lattice QCD computations of
the UKQCD group.

The origin of this disagreement has been investigated.
Starting from  a broken VMD based Lagrangian, we have shown how to 
deduce the (WZW) $\eta/\eta'$ two--photon amplitudes on the one hand, and 
the expectation values $\langle0|J^{0,8}|\eta/\eta'\rangle$ 
on the other hand
($J$'s are the axial currents), which gives the customary
(ChPT) definition of decay constants. 

The disagreement reported above has been
traced back to inconsistent definitions for the same parameters
provided by ChPT and the WZW Lagrangian (through the former definitions 
of Current Algebra). This inconsistency is a pure consequence
of breaking SU(3) symmetry. For instance, it was shown
that none of the two angles of extended ChPT can appear {\it as such} 
in the two--photon 
decay amplitudes. In some way, besides the angles $\theta_8$ and $\theta_0$
recently introduced by Kaiser and Leutwyler, the standard angle
$\theta_P$, which still describes the $\eta/\eta'$ wave--function mixing, 
goes on playing an important role, the main one in radiative decays. 
It has been shown that $\theta_P \simeq \theta_8/2$, 
within a ChPT motivated Lagrangian framework,  perfectly
accounting for all data on radiative decays and all accessible ChPT expectations.
Moreover, analysis of numerical correlations has illustrated
why the use of $\theta_P$ is more appropriate in order
to get {\it accurate} measurements of physics parameters,
including the standard ChPT ones.
 
\vspace{1.cm} 
 
We thus have clearly illustrated that, the HLS--FKTUY model,
supplemented with SU(3) flavor breaking \`a la BKY and
nonet symmetry breaking, as was introduced in Ref. \cite{rad}
was equivalent to the ChPT  Lagrangian ${\cal L}^{(0)}+{\cal L}^{(1)}$.
Besides the successful  description of all radiative decays, broken VMD 
thus meets all the requirements of extended ChPT (the two mixing angles and 
the two decay
constants) with good accuracy. It was then shown that the relevant (and
self--consistent) angle pattern is $\theta_0 \simeq 0^{\circ}$, 
$\theta_8 \simeq -20^{\circ}$ and $\theta_P \simeq -10^{\circ}$. 
Correspondingly, we have simultaneously $\overline{f_8}=0.82 f_{\pi}$
when using the WZW (or Current Algebra) definition and
$f_8=1.33 f_{\pi}$, when using standard ChPT definition.
Subsequently, we have shown that the $\overline{f}_i$ cannot
be interpreted as isoscalar meson decay constants because
of flavor SU(3) breakdown. Therefore, VMD phenomenology
is indeed able to provide ChPT with quite reliable input.
With this respect, an important feedback from ChPT would be a more 
precise estimate of $\theta_0$ and/or a (reliable) theoretical error. 

This study has led us to several additional conclusions:

{\bf i/} Because  $\theta_0 = 0$ is observed with
a quite impressive precision, it has been possible to
relate   nonet symmetry breaking (the parameter $x$) and 
the mixing angle $\theta_P$. To our knowledge such a relation has never 
been reported. 

{\bf ii/} The nonet symmetry breaking parameter $\lambda$
which weights the additional singlet contribution to the Lagrangian
is small ($\lambda \simeq 0.20 \pm 0.04$). It coincides
with the usual OZI breaking parameter $\Lambda_1$ of ChPT.

{\bf iii/} As consequence of 
$x= 0.901$, it has been shown, that no glue component is needed
inside the $\eta'$ meson. 

{\bf iv/} By relating $x$ and $\theta_P$,
the condition $\theta_0=0$ allows to account for
observed correlations in fitting radiative decays.
Additionally, this leads us to propose a 4--parameter model
to account for all data on radiative decays (including $K^{*\pm} \ra K^{\pm} \gamma$);
this is by far the most constrained model ever proposed and we proved
that it is quite successful.

 {\bf v/} The quark mass ratio deduced from VMD information 
is $m_s/\hat{m}= 21.2 \pm 2.4$.

{\bf vi/} Departure from the classical Gell--Mann--Okubo
quadratic mass relation is observed at only the percent level.

\vspace{1.0cm}
\begin{center}
{\bf Acknowledgements}
\end{center}
We thank B. Moussalam (IPN, Orsay), who first remarked
that the disagreement with ChPT expectations reported
in Ref. \cite{rad} could simply follow from having defined 
the decay constants from Current Algebra relations. 
We are also deeply indebted to Victor L. Chernyak (Budker Institute,
Novosibirsk), for his patience in many illuminating discussions
about the effects of symmetry breaking in the conceptual
framework of ChPT, which led us to formulate
a framework which can encompass both the phenomenology
of radiative decays and all current expectations of ChPT.
We finally acknowledge useful correspondence and discussions
with Th. Feldmann (Wuppertal University) about the connection
of our modeling with usual concepts of ChPT.
HOC was supported by the US Department of Energy under contract
DE--AC03--76SF00515. 

\newpage
 
\begin{center}
\begin{tabular}{|| c  | c  | c | c ||}
\hline
\hline
\hhhc  \hhhb  Process     &  $x$ and $\theta_P$  &  $x$ and $\theta_P$ &  PDG \\
\hhhc  \hhhb              & Related    &  Unrelated & \\
\hline
\hline
$\rho \rightarrow \pi^0 \gamma$     $(\times 10^4)$ & $5.16 \pm 0.03$ & $5.16 \pm 0.03$ & $6.8 \pm 1.7$ \hhhu\\
\hline
$\rho \rightarrow \pi^\pm \gamma$   $(\times 10^4)$ & $5.12 \pm 0.03$ & $5.12 \pm 0.03$ & $4.5 \pm 0.5$ \hhhu \\
\hline
\hline
$\rho \rightarrow  \eta \gamma$     $(\times 10^4)$ \hhhu & $3.16 \pm 0.05$  & $3.19 \pm 0.10$ & $2.4^{+0.8}_{-0.9}$\\
\hline
$\eta' \rightarrow \rho \gamma$     $(\times 10^2)$ \hhhu & $33.3 \pm 1.26$ & $34.5 \pm 2.1$  & $30.2 \pm 1.3$ \\
\hline
\hline
$K^{*\pm} \rightarrow K^\pm \gamma$ $(\times 10^4)$ \hhhu & $9.80 \pm 0.94$  & $9.80 \pm 0.93$ & $9.9 \pm 0.9$ \\
\hline
$K^{*0} \rightarrow K^0 \gamma$     $(\times 10^3)$ \hhhu & $2.32 \pm 0.02$  & $2.32 \pm 0.02$ & $2.3 \pm 0.2$\\
\hline
\hline
$\omega \rightarrow \pi^0 \gamma$   $(\times 10^2)$ \hhhu & $8.49 \pm 0.05$  & $8.49 \pm 0.05$ & $8.5 \pm 0.5$ \\
\hline
$\omega \rightarrow \eta \gamma$    $(\times 10^4)$ \hhhu & $7.81 \pm 0.11$  & $7.88 \pm 0.23$ & $6.5 \pm 1.0$\\
\hline
$\eta' \rightarrow \omega \gamma$   $(\times 10^2)$ \hhhu & $2.83 \pm 0.11$  & $2.94 \pm 0.19$ & $3.01 \pm 0.30$ \\
\hline
\hline
$\phi \rightarrow \pi^0 \gamma$     $(\times 10^3)$ \hhhu & $1.28 \pm 0.12$  & $1.27 \pm 0.12$ & $1.31 \pm 0.13$\\
\hline
$\phi \rightarrow \eta \gamma$      $(\times 10^2)$ \hhhu & $1.28 \pm 0.02$  & $1.27 \pm 0.04$ & $1.26 \pm 0.06$ \\
\hline
$\phi \rightarrow \eta' \gamma$     $(\times 10^4)$ \hhhu & $0.59 \pm 0.02$  & $0.60 \pm 0.03$ & $1.2^{+0.7}_{-0.5}$\\
\hline
\hline
$\eta \rightarrow \gamma \gamma$    $(\times 10^2)$ \hhhu & $38.87 \pm 0.75$ & $39.3 \pm 1.8$  & $39.21 \pm 0.34$ \\
\hline
$\eta' \rightarrow \gamma \gamma$   $(\times 10^2)$ \hhhu & $2.09 \pm 0.08$  & $2.17 \pm 0.10$ & $2.11 \pm 0.13$ \\
\hline
\hline
\end{tabular}

\parbox[t]{16.0cm}{ 
      {\bf Table 2 :}  
Radiative decay branching fractions. The first two data columns display the fit results 
using the $K^*$ model of Refs. \cite{rad,conf}; in the first data column (present work)
the $x$--$\theta_P$ of Eq. (\ref{chpt8}) is switched on while, in the second one, it is not.
The data for $\eta \ra \gam$ is the recommended value \cite{PDG98}. The last data column 
displays the accepted values from the 1998  Review of Particle Properties \cite{PDG98}.
}
\end{center}

\appendix
\section{The anomalous decay term}

We shall now briefly discuss the equivalence of the anomalous
decay amplitude $T(P\ra \gamma\gamma)$ as calculated either from
the divergence of the axial current, or the WZW Lagrangian \cite{victor}.
Let us recall the form of the axial current given in Eq.(\ref{nsk2}).
We then take the divergence of this to obtain{\footnote{
 In this Section we use for conciseness the notation
$P=\sum_{a=0, \cdots 8} P^a T^a$; thus for instance $P^0=\eta_0$. 
}}
\be
\pa^\mu J_\mu^a = 2f{\rm Tr}[T^aX_AT^bX_A]\pa^2 P^b+\lambda f\delta^{a0}
\pa^2 P^0.\label{dJ}
\ee
Now let us turn to the equations of motion for the pseudoscalar fields,
namely,
\be
\pa_\mu\frac{\pa{\cal L}}{\pa (\pa_\mu P)}-\frac{\pa{\cal L}}{\pa P}=0,
\ee
leading to (allowing for a pseudoscalar mass term)
\be
2{\rm Tr}[T^aX_AT^bX_A]\pa^2 P^b+\lambda \delta^{a0}
\pa^2 P^0=m_P^2 P^a -\frac{C}{f}
\varepsilon^{\mu\nu\alpha\beta}F_{\mu\nu}F_{\alpha\beta}
{\rm Tr}[Q^2T^a],\label{eqmot}
\ee
where $C$ is a well known dimensionless constant \cite{WZW2}.

We are now in a position to show
that the amplitude obtained form the axial current is equivalent
to that obtained from the anomalous Lagrangian.
First let us consider 
$\langle A A|\pa^\mu J_\mu^a|0\rangle$. As a total divergence this
vanishes, in accordance with the fact that there are no truly
massless particles in the spectrum (we are not considering the chiral limit).
So using this along with Eqs.~(\ref{dJ}) and (\ref{eqmot}) we have
\be
fm^2_P\langle A A|P^a|0\rangle=C\langle A A|F_{\mu\nu}\tilde{F}^{\mu\nu}
{\rm Tr}[Q^2T^a]|0\rangle.
\ee
As $q\ra 0$ 
$\langle X|P|0\rangle = \langle X|P\rangle$,
hence, as $m^2_P$ is absorbed in the definition of the amplitude, we have
\be
T(P^a\ra AA)
=\frac{C}{f}\langle A A|F_{\mu\nu}\tilde{F}^{\mu\nu}
{\rm Tr}[Q^2T^a]|0\rangle.
\ee
We see that the result we have obtained, starting with the axial
current, is the same as one would obtain from the anomalous Lagrangian
term. {This equivalence extends to $VVP$ interactions.}


\newpage
\begin{thebibliography}{99}

\bb{rad} 
M.~Benayoun, L.~DelBuono, S.~Eidelman, V.~N.~Ivanchenko and H.~B.~O'Connell,
Phys.\ Rev.\  {\bf D59}, 114027 (1999)
[hep-ph/9902326].


\bb{HLS} M. Bando, T. Kugo, S. Uehara, K. Yamawaki and T.
Yanagida, Phys. Rev. Lett. {\bf 54}, 1215 (1985);
M. Bando, T. Kugo and K. Yamawaki, Phys. Rep. {\bf 164}, 217 (1985).

\bibitem{Seiberg:1995pq}
N.~Seiberg,
Nucl.\ Phys.\  {\bf B435}, 129 (1995)
[hep-th/9411149].


\bb{heath} M. Benayoun and H.B. O'Connell, Phys. Rev. {\bf D58} (1998) 074006.


\bibitem{Harada:1999zj}
M.~Harada and K.~Yamawaki,
Phys.\ Rev.\ Lett.\  {\bf 83}, 3374 (1999)
[hep-ph/9906445].


\bb{FKTUY} 
T.~Fujiwara, T.~Kugo, H.~Terao, S.~Uehara and K.~Yamawaki,
Prog.\ Theor.\ Phys.\  {\bf 73}, 926 (1985).


\bb{BKY} M. Bando, T. Kugo and K. Yamawaki, Nucl. Phys. {\bf B259}, 
493 (1985).


\bb{conf} M.~Benayoun {\it et al.}, Proceeding of
the international Workshop ``$e^+e^-$ Collisions from $\phi$ to $J/\psi$'',
Novosibirsk, March 1--5, 1999, to be published, hep-ph/9906372.

\bb{mixing} 
M.~Benayoun, L.~DelBuono, P.~Leruste and H.~B.~O'Connell,
nucl-th/0004005, to be published in Eur. Phys. Journ. C.


\bb{PDG98} C. Caso {\it et al.} Eur. Phys. Journ. {\bf C3} (1998) 1.



\bb{morpurgo} G. Morpurgo, Phys. Rev. {\bf D 42} (1990) 1497.

\bb{odonnel} P. O'Donnell, Rev. of Mod. Phys. {\bf 53} (1981) 673.


\bb{DHL}  J. F. Donoghue, B. R. Holstein and Y. R. Lin, Phys. Rev.
Lett. {\bf 55} (1985) 2766.

\bb{GILMAN} F. J. Gilman and R. Kauffman, Phys. Rev. {\bf D36} (1987) 2761.

\bb{ball} P. Ball, J.--M. Fr\`ere and M. Tytgat, 
Phys. Lett. {\bf B365} (1996) 367.

\bb{veneziano}  G.M. Shore and G. Veneziano, Nucl. Phys. {\bf B381} (1992) 3.

\bb{basu1} S. Basu and B. Bagchi, Zeit. f\"ur Phys. {\bf C 37} (1987) 69.


\bb{basu2} S. Basu, B. Bagchi and A. Lahiri  Phys. Rev. {\bf D 37} (1988) 1250.



\bb{leutw} H. Leutwyler, Nucl. Phys. Proc. Suppl. {\bf 64}, 223 (1998)
[hep-ph/9709408].

\bb{leutwb} R. Kaiser and H. Leutwyler, hep-ph/9806336.

\bb{feldmann}
T.~Feldmann,
Int.\ J.\ Mod.\ Phys.\  {\bf A15}, 159 (2000)
[hep-ph/9907491].

\bb{kroll1} T. Feldmann and P. Kroll, Eur. Phys. J. {\bf C5} (1998) 327;
Phys. Rev. {\bf D58} (1998) 057501.

\bb{kroll2} T. Feldmann, P. Kroll and B.Stech,
Phys.\ Lett.\  {\bf B449} (1999) 339
 [hep-ph/9812269].

\bb{shore} 
G.~M.~Shore,
Nucl.\ Phys.\  {\bf B569}, 107 (2000)
[hep-ph/9908217].


\bb{WZW1} J. Wess and B. Zumino, Phys. Lett. {\bf 37B} (1971) 95.

\bb{WZW2} E.~Witten, Nucl. Phys. {\bf B223} (1983) 422.

\bb{ukqcd}
C.~McNeile and C.~Michael  [UKQCD Collaboration],
hep-lat/0006020.

\bb{scadron1} 
A.~Bramon and M.~D.~Scadron,
Phys.\ Lett.\  {\bf B234}, 346 (1990).

\bb{scadron1b} 
A.~Bramon, R.~Escribano and M.~D.~Scadron,
Eur.\ Phys.\ J.\  {\bf C7} (1999) 271
[hep-ph/9711229].

\bb{scadron2} 
D.~Kekez, D.~Klabu\v{c}ar and M.~D.~Scadron,
hep-ph/0003234.

\bb{kekez} D. Klabu\v{c}ar and D. Kekez, Phys. Rev. {\bf D 58} (1998) 096003
[hep-ph/9710206].

\bb{scadron3} R. Delburgo, D. Liu and M.D. Scadron, 
Int. J. Mod. Phys. {\bf A 14} (1999) 4331
[hep--ph/9905501].

\bb{gedalin}
E.~Gedalin, A.~Moalem and L.~Razdolskaya,
nucl-th/0006073.

\bb{GL85}J. Gasser and H. Leutwyler, Nucl. Phys. {\bf B250} (1985) 465;
Ann. Phys. (NY) {\bf 158} (1984) 142.

\bb{DGH} J.F. Donoghue, E. Golowich, B.R. Holstein,
``Dynamics of the Standard Model'', Cambridge University Press,
Cambridge  UK, 1996, P. 204.

\bb{pich1} A. Pich, ``Effective Field Theory'', hep-ph/9806303.


\bb{pich2} 
A.~Pich,
Rept.\ Prog.\ Phys.\  {\bf 58}, 563 (1995)
[hep-ph/9502366].

\bb{ben2} M. Benayoun,  Ph.~Leruste, L.~Montanet and J.--L.~Narjoux,
Zeit. Phys. {\bf C65} (1995) 399.

\bb{kou} 
E.~Kou,
hep-ph/9908214.

\bb{ben4} 
M.~Benayoun, S.~Eidelman, K.~Maltman, 
H.~B.~O'Connell, B.~Shwartz and A.~G.~Williams,
Eur.\ Phys.\ J.\  {\bf C2}, 269 (1998)
[hep-ph/9707509].

\bb{cmdpi} 
 R.~R.~Akhmetshin {\it et al.}  [CMD-2 Collaboration],
``Measurement of $e^+ e^- \rightarrow \pi^+ \pi^-$
cross-section with CMD-2 around $\rho$  meson,''
hep-ex/9904027.



\bb{BGP} A.~Bramon, A.~Grau and G.~Pancheri, 
Phys. Lett. {\bf B344}, 240 (1995).

\bibitem{rod}
R.J.~Crewther,
Riv. Nuovo Cim. {\bf 2}, 63 (1979).

\bibitem{christos}
G.A.~Christos,
Phys. Rept. {\bf 116}, 251 (1984).

\bibitem{tH}
G.~'t Hooft,
Phys. Rept. {\bf 142}, 357 (1986).
 
\bb{victor} V.L.~Chernyak, private communication.

\bb{review} H.B.~O'Connell, B.C.~Pearce, A.W.~Thomas and A.G.~Williams,
	Prog. Part. Nucl. Phys. {\bf 39} (1996) 201.

\bb{abegg} R. Abegg {\it et al.} Phys. Rev. {\bf D 53} (1996) 11.

\bb{chan} M.S. Chanowitz, Phys. Rev. Lett. {\bf 35} (1975) 977;
Phys. Rev. Lett. {\bf 44} (1980) 59.

\bb{signal} F.~Cao and A.~I.~Signal,
Phys.\ Rev.\  {\bf D60} (1999) 114012
[hep-ph/9908481].


\bb{holstein} E.P. Venugopal and B.R. Holstein, Phys. Rev. {\bf D 57}
(1998) 4397. 

\bb{purlatz} T. Purlatz {\it al.},   Proceeding of
the international Workshop ``$e^+e^-$ Collisions from $\phi$ to $J/\psi$'',
Novosibirsk, March 1--5, 1999, to be published.

\bb{phietp2} V. M. Aulchenko {\it et al.}, 
Phys. Lett. {\bf B436} (1998) 199; 
M.N. Achasov  {\it et al.}, Preprint Budker INP 98--65, Novosibirsk, 1998. 

\bb{phietp1}  R.R. Akhmetshin  {\it et al.}, Phys. Lett. {\bf B415} (1997) 445.

\end{thebibliography}
\end{document}